**Molecular architecture of TE-seeded chromosome-naïve initiation networks of protein-protein interactions delineates the Human Protein Map of the assembly of multiprotein complexes.**

Gennadi V. Glinsky[1]

[1] Institute of Engineering in Medicine
University of California, San Diego
9500 Gilman Dr. MC 0435
La Jolla, CA 92093-0435, USA
Correspondence:
genlighttech@gmail.com

**Abstract**

Upon binding to transposable elements (TE) – seeded DNA templates, initiator transcription factors (TFs) are poised to activate the hybrid assembly pathways (HAP) of endogenous multiprotein complexes (EMC) by iteratively engaging protein-protein interaction (PPI) cascade chain reactions and creating cell type-specific pools of protein dimers. In this contribution, molecular features and inferred biological functions of the constituents of PPI cascades were elucidated within the context of cell type-specific intracellular protein monomer pools documented in the Human Protein Atlas (HPA) of 154 human cell types. Present analyses revealed the following key features of the molecular architecture and functional performance of the 4 families of PPI cascade constituents within the human body-wide contexts of intracellular proteomes: a) patterns of ubiquitous expression and balanced representation; c) the universal availability and engagements; d) constitutive {invariant) and cell type-specific assembly profiles of EMCs. The biologically meaningful behavior of the PPI cascades is demonstrated by the cell-type-resolved multiprotein complex-engagement analysis across the HPA of cellular diversity catalogue of human body, which faithfully recovers without supervision the intrinsic lineage structures of EMCs and lineage-specific master-regulator multiprotein assemblies. Overall, the PPI cascade constituents have the capacity to generate 51,830 protein dimers within the intracellular proteome availability contexts of 154 types of human cells of distinct specializations. Present analyses revealed critical impacts of the intracellular persistence of proteinopathy proteins (APP, HTT, SNCA, MAPT) that may cause the damage to the PPI cascade performance by sequestering 44 synaptic proteins. The sequestration of synaptic proteins withdraws crucial structural-functional components of synapse integrity suggesting a mechanism contributing to the pathogenesis of neurodegeneration. Conversely, reconstitution of availability of defined synaptic structural proteins appears to restore stoichiometry and recovers the lost assembly of synaptic protein dimers.

## Introduction

Recent reports documented multiple lines of observations revealing the initiation mechanism of protein-protein interactions (PPI) of hybrid assembly pathways (HAP) of endogenous multiprotein complexes, EMC (Glinsky, 2026a; b). According to HAP concept, chromosome-naïve networks of DNA sequences derived from transposable elements (TE) function as initiation templates of PPI chain reactions which continually creates cell type-specific pools of hetero- and homodimers. Upon release from DNA templates, protein dimerization products serve as on-demand attractors of specific sets of monomers for the EMC assembly completion. Identification and initial characterization of key constituents of HAP of EMC (Glinsky, 2026 a; b) revealed a set of ~700 proteins designated initiator transcription factors (TF). Initiator TFs bind to double-stranded DNA templates in a sequence-specific manner. Upon binding to transposable element-seeded DNA templates, initiator TFs are poised to initiate PPI cascade chain reactions creating cell type-specific pools of protein homo- and heterodimers (Glinsky, 2026 a; b).

Follow-up differential Gene Set Enrichment Analyses (GSEA) demonstrated that this set of initiator TFs could be further divided into 2 families based on distinct patterns of their expression in human brain regions: 1) a set of 545 proteins manifesting region-specific profiles of significantly enriched or depleted expression in human brain; 2) a set of 171 proteins expression of which is detectable in hundreds of brain regions with no evidence of region-specific expression enrichments or depletion (Glinsky, 2026 a;b). However, independent GSEA of 545 and 171 initiator TFs revealed numerous examples of identical significantly enriched records of biological and disease pathways (11 complementation categories of GSEA employing the WikiPathways 2024 Human database); Mammalian Phenotypes (104 complementation categories of GSEA employing the MGI Mammalian Phenotype Level 4 2024 database); phenotypes of human diseases (37 complementation categories of GSEA employing the DisGeNET database). Correspondingly, these 2 families of HAP constituents were designated as the principal initiator TFs and complementary initiator TFs (detailed summaries of these findings is available in the Supplementary Table S1).

Analyses of PPI cascade patterns of initiator TFs demonstrated that HAP-associated PPI chain reactions on DNA templates could concurrently proceed by branching into 2 modes of operations: TF-TFs PPI cascade branch and TF-HUB proteins PPI cascade branch (Glinsky, 2026 b). It has been observed that the sequence of PPI cascade chain reactions reach the PPI cascade equilibrium at saturation levels of PPI constituents of 245 TFs and 351 TF-HUBs of TF-TFs PPI cascade branch and TF-HUB proteins cascade branch, respectively (Glinsky, 2026 b). Correspondingly, these observations identified 2 mechanistically distinct families of HAP constituents: 1) a set of proteins comprising the equilibrium (plateau) level constituents of the TF-TFs PPI cascade branch; 2) a set of proteins comprising the equilibrium (plateau) level constituents of the TF-HUB proteins PPI cascade branch. Therefore, previous reports indicate that the canonical panel of proteins engaged in PPI cascade chain reactions during the execution of HAP on DNA templates consists of 4 distinct families of HAP constituents: 1) principal initiator TFs; 2) complementary initiator TFs; 3) an equilibrium (plateau) stage PPI TF-TFs binders; 4) an equilibrium (plateau) stage PPI TF-HUB proteins binders. However, the genetic identities and numerical values assigned to each family of HAP constituents were based on genome-wide analyses of TFBS and PPI cascade chain reactions associated with LTR5_Hs loci and detailed analyses of these families of HAP constituents in the contexts of diverse patterns of protein availabilities in human cells have not been reported.

Multiprotein complexes are the essential structural and functional units of all vital cellular processes, and the PPI between monomer subunits that build them are catalogued in interaction databases that are largely cell-type agnostic. A curated complex such as those in the CORUM database, or PPI terms in a hub-protein library, record sets of proteins that can associate — not talking into account whether any particular cell contains all of them or a fraction of them at the same time. This gap between the interactome as catalogued and the interactome that may contribute to the EMC assembly is the practical problem behind most network-based inference in contemporary cell and developmental biology. A method that reasons over the full catalogue will predict the same complexes everywhere, and a method that reasons over what a cell expresses has to specify how expression constrains assembly.

Iterated enrichment offers one way to bridge the gap. Beginning from a seed set of proteins, one tests every term in an interaction library for over-representation of seeds among its members, adds the members of the significant terms to the seed set, and repeats until the set stops changing. The fixed point — the plateau — is a self-consistent set of proteins that the library links to the seeds through statistically supported steps. This construction has been previously applied to a panel of 716 transcription factors whose binding sites are encoded by transposable elements (Glinsky, 2026 a; b), and reported observations indicate that the panel converges on plateaus of 245 proteins in a transcription-factor interaction library and 351 in a hub-protein PPI library (Glinsky, 2026 b). One of the interpretations that might be derived from this model is that these PPI cascade plateaus may describe PPI networks that the initiator panel is capable of nucleating.

Along the line of this interpretation is an assumption that has not been tested: that the PPI cascade plateau is a property of the seed panel. An iterated enrichment procedure run on a fixed library is a dynamic system with its own attractors, and if the library's attractor basin is wide enough, any sufficiently large or sufficiently hub-rich seed set will converge to the same plateau. Should that be so, a plateau observed for a particular panel would carry no information about that panel, and the biological content attributed to it would reflect the library content. Overall, the critical questions whether the reported PPI cascade performance is capable to discriminate between the different cell types and whether this PPI cascade-driven discrimination is aligned with cellular specialization states remain open.

A second untested assumption concerns the availability of pools of proteins that are engaged in PPI in a defined intracellular environment. The operation of reported PPI cascade was tested on the full library universe, in which every catalogued protein is available at every step. Real cells express a subset of the proteome, and single-cell consensus expression atlases now make that subset measurable at the resolution of individual cell types across whole organisms. If the cascade is restricted at every step to the proteins a given cell expresses — both in which library terms may be used and in the statistical background against which enrichment is judged — then its output becomes a per-cell-type quantity. In this context, the PPI cascade plateau becomes a report about what a particular cell can assemble rather than about what the interactome contains. Whether that restriction produces a map with biologically-meaningful cell-type structure, or merely rescales a common message, is an empirical question that should be tested.

In this study both assumptions have been tested and the results were used to build an atlas of 154 human cell types. The independent computational analyses of the reported PPI cascade constituents and their performance were carried-out by re-computing them against size-matched, degree-matched and transcription-factor-census null seed sets, and by asking whether the two initiator subfamilies build separable networks. Next, the PPI cascade memberships were restricted to the measured proteome of each of 154 human cell types of the Human Protein Atlas and scored the resulting engaged protein sets against 1,127 curated multiprotein complexes. These analytical experiments were executed under two independent definitions of protein availability, so that transcriptome breadth can be separated from cell identity as a source of apparent variation. The resulting multiprotein complex-engagement map was characterized by its effective dimensionality, and the PPI cascade performance was evaluated by assessing how quickly it reaches its fixed point. In this contribution, molecular features and inferred biological properties of the PPI cascades governed by a canonical set of 1,122 HAP constituents were elucidated within the contexts of intracellular protein pools available in the Human Protein Atlas (HPA) of 154 distinct cell types. Results of present analyses documented several key features of the molecular architecture and functional performance of the PPI cascade constituents comprising of: a) the ubiquitous expression pattern; b) the universal representation of 4 families of PPI cascade constituents; c) the nucleator functions of a sub-set of principal initiator TFs; d) both constitutive {invariant) and cell type-specific assembly profiles of endogenous multiprotein complexes in defined intracellular proteome contexts. Lastly, a PPI cascade sequestration hypothesis of proteinopathy pathologies has been formulated and computationally validated on a large panel of 154 human cell types.

## Results

### Independent re-computation of the initiator-seeded PPI cascade: null behaviour, family structure, and assembly in defined cellular proteomes

Previous studies of the initiator TF-seeded PPI cascades analyzed the canonical set of HAP constituents defined by the PPI patterns of 716 proteins having sequence-specific binding sites within genomic loci encoded by transposable elements (Glinsky, 2026 a; b). One of the shortcomings of the previously reported analyses was the limited considerations of the impact of the availability of proteins in a cell on the PPI cascade performance. Both reported branches of the PPI cascade reproduce as proper subsets of the published lists. The TF–TF cascade saturates at 207 proteins by iteration 6 and the HUB cascade at 346 proteins by iteration 10 (Supplementary Text ST1). All 207 and all 346 are in the published 245-member and 351-member lists; the reciprocal count — proteins reached here but absent from the published lists — is zero in both branches. That asymmetry is important since a re-computation that disagreed on mechanism would produce members the published PPI cascade never reached; this one produces none. The 38 TF–TF members not identified during the re-computation occupy ranks 184–245 of the published ordering of 245 TFs (median 224.5), with none in the top 100 and 32 in ranks 201–245. The 5 HUB members not identified during the re-computation (HMT1, MDFI, NAF1, RPOB, UBC7) sit at ranks 345 and 348–351 of the published ordering of 351 HUB members (Supplementary Text ST1). The cascade therefore walks down the published ranking and stops just short of the bottom, which is the signature of a stringency difference, likely reflecting a slightly different BH cutoff or a different Enrichr library snapshot.

The union of the initiator panel and the two cascade branches is 1,123 entries, of which 1,122 are named genes — the cumulative network size of HAP constituents reported in elsewhere (Glinsky, 2026 a; b). The HUB branch loses 106 members in the cellular arm because the Enrichr HUB library carries yeast entries (SSA1, SSB1/2, HSP82, HSC82, TDH3, YDJ1, SIS1, KAR2, RVB1/2, TUB1/2, ACT1 and others; Supplementary Summary S1) that have no human single-cell measurement and so cannot enter a dimer prediction. The TF–TF branch has full atlas coverage.

Random seed sets drawn from the library background reach nothing: the cascade dies at zero in both branches ($p = 0.016$). Once the seeds are hub-aware, the picture inverts (Table 2). Degree-matched random seeds of the same size reach 207–208 proteins in the TF–TF library (mean 207.3, SD 0.5) and up to 347 in the HUB library, and the membership they reach is the same membership: mean Jaccard 0.999 and maximum 1.00 against the observed plateau. Seeds drawn at random from the human TF census (GO:1990837) reproduce the observed plateau exactly, Jaccard 1.00 in both branches.

These observations are in agreement with the conclusion that 207- and 346-protein equilibrium plateaus are attractors of the interaction libraries under iterated enrichment; any comparably sized, comparably hub-heavy seed set lands on them. The specificity of the initiator panel, if it is to be observed, has to be demonstrated at small seed number or in a restricted proteome — not at the plateau.

The 545 principal initiators (see Introduction and the Supplementary Table S1 for definitions) and the 171 complementary initiators partition the panel exactly: all 545 are members of the 716-protein panel, the remainder is exactly the 171 reported initiator TFs, and the two sets are disjoint and complete.
Analyzed for the PPI cascade performance separately, they behave identically. In the TF–TF library, 413 mapped principal seeds and 80 mapped complementary seeds both reach 208 proteins by iteration 7, against a size-matched panel-subset null of 207–208 ($p = 0.645$ and $p = 0.871$). In the HUB library, 375 and 73 mapped seeds both reach 346 by iteration 10, with $p = 1$ for both (Supplementary Text ST1). The two family plateaus overlap at Jaccard 1.00 in both branches, with zero principal-private and zero complementary-private members (Supplementary Table S2). The interpretation is not that the families are biologically equivalent. Distinctions in brain-region enrichment profiles are based on an expression phenotype, and the PPI cascade as computed here has no expression term in it. The interpretation is that the PPI cascade topology cannot highlight the difference. Therefore, any claim of functional distinctions between the families has to come from a model that carries the expression information (see below).

**In a defined cellular proteome the PPI cascade assembly profiles become cell type-specific**

In the next series of analytical experiments, the interrogated constituents of the PPI cascade members were restricted to the proteins that are expressed in a specific cell type (HPA single-cell consensus, nTPM ≥ 1). Results of these analyses revealed the changes of the PPI cascade performance defined by both quantitative and qualitative metrics. The initial experiments were carried out on brain excitatory neurons and hepatocytes (Table 1).

In brain excitatory neurons, 7,997 proteins are available, including 696 of the 1,122 network members and 421 of the 716 initiators; 300 initiators map into the restricted TF–TF library and 290 into the restricted HUB library. The cascade reaches 150 proteins in the TF–TF branch (85 → 136 → 143 → 147 → 150 → 150) and 216 in the HUB branch (60 → 174 → 212 → 216 → 216), engaging 339 proteins in total. That supports 171 homodimer-capable proteins, 5,923 mutually engaged heterodimers, and 7,694 dimers inside the 1,122-protein panel. Scored against CORUM, 257 complexes are enriched among the cascade products at q < 0.05 and 28 have every subunit engaged.

In hepatocytes, 6,888 proteins are available (612 network members, 348 initiators). The plateaus are close in size — 153 in the TF–TF branch, 211 in the HUB branch, 337 proteins engaged — and the dimer count is comparable at 6,000 mutually engaged heterodimers, but the complexes that assemble are different: 225 enriched at q < 0.05, 25 fully engaged.

The complex-level contrast is the substantive finding of this analysis (Figure 1). Cell-cycle and growth-signaling assemblies are engaged in hepatocytes and not in neurons — the CDC2–CCNA2–CDK2 complex goes from 0.00 engaged in neurons to 0.67 in hepatocytes, NOTCH1–P56LCK–PI3K from 0.33 to 1.00, cell-cycle kinase complex CDK2 from 0.25 to 0.75 — while chromatin, DNA-damage and neuronal assemblies run the other way: the BMI1–MEL18–MPH1–M33 polycomb repressor complex 0.75 in neurons versus 0.25 in hepatocytes, and MKK4–ARRB2–JNK3, BRCA1–BARD1–POLR2A, E2F4–p107–cyclin E and the RB–NEUROD1–NGFI-B complex each 0.67 versus 0.33. A small set of complexes is fully engaged in both cells and forms the cell-invariant core of the prediction: P300–MDM2–P53 (engaged fraction 1.00 in both; atlas mean 0.996 across the 154 cell types, ranking 11th in neurons and 64th in hepatocytes) and PU.1–SIN3A–HDAC (1.00 in both; 7th in neurons, 114th in hepatocytes — invariant in engaged fraction while still well separated in atlas rank).

Notably, the TF–TF library yields zero homodimers in either cell type, because its terms do not include the TF itself; every predicted homodimer comes from the HUB library's self-membership. The model performs in a real cellular context: given a measured proteome it produces a specific, testable inventory of dimers and complexes, and that inventory differs between cell types in the direction their biology predicts (Figure 1). Above analyses have several limitations within the contexts of the intracellular PPI cascade performance. Significantly, the nTPM ≥ 1 cellular availability call is a threshold on mRNA, not a direct protein measurement. Thus, mRNA-inferred availability is the weakest link in the dimer predictions. Dimer calls are library co-membership supported by the reported PPI between proteins. These predicted protein dimers do not reflect the structural or affinity predictions within the living cells intracellular contexts, and the limiting-capacity ranking (min nTPM of the two partners) could be viewed as a stoichiometric proxy, not a directly measured abundance of the complex or its components.

**Universal availability and engagement of the 4 families of PPI cascade constituents in 154 human cell types**

Assessments of availability and engagement metrics were carried-out based on expression analyses data for 154 individual cell types and results were plotted for visualization as the availability (Figure 2) and the engagement (Figure 3) profiles of the 4 families of PPI cascade constituents. To ensure that observations are not artefacts of the arbitrary selected expression threshold, profiles of availabilities and engagements of the PPI cascade constituents in each individual cell type of the HPA panel were evaluated at the several cut-off levels of the expression threshold (Figures 2 and 3). Several key features of the molecular architecture of the 4 families of PPI cascade constituents were ascertained during the analyses of their expression and engagements within the cell type-specific contexts of intracellular human proteomes.

Every family of the PPI cascade constituents manifests ubiquitous expression and balanced representation of members of all 4 families in human cells (Figure 2). A total of 42/544 (7.7%) measured principal initiator TFs (P545) and 9/166 (5.4%) measured complementary initiator TFs (C171) are expressed in all 154 cell types. Apparently larger fractions of members of the PPI cascade TF-TF branch (T245) and TF-HUB protein branch (H350) manifest similarly ubiquitous expression profiles: expression of 20% of TF-TF branch constituents and 40% of TF-HUB protein branch constituents was detected in all 154 cell types (Figure 2). Notably, the rule of the ubiquitous expression is not an artefact of one arbitrary selected threshold and it appears universally applicable to 144 somatic differentiated cell types (Figure 2).

Next, the metrics of the availability and engagements of the PPI cascade constituents were evaluated, considering the expression level in a specific cell type as the proxy for member’s availability in a cell (Figure 2) and the inferred ability to form protein dimers during the PPI cascade performance as the proxy for the member’s engagement (Figure 3). Results of these analyses performed at different expression threshold levels documented the universal availability and engagements of members of 4 families of PPI cascade constituents, while members of the TF-HUB protein branch of the PPI cascade appear to manifest higher engagement levels (Figures 2 and 3).

**Cell-type-resolved expression of PPI cascade constituents and engagement of endogenous complexes across 154 human cell types**

The cascade model treats assembly as availability-gated: an initiator can nucleate a complex only where its partners are co-present in a cell. To test whether this produces the cell type-specific structure rather than a uniform readout, we projected 1,127 CORUM core complexes onto the single-cell transcriptomes of 154 human cell types (Human Protein Atlas), scoring each complex in each cell type as the fraction of its subunits drawn from that cell's available proteome. Under the expression rule (nTPM ≥ 1), an average of 7,388 proteins were available per cell type (SD 519, range 5,092–8,327), of which 392 (SD 41.9) were initiators and 664 (SD 51.7) were network members. Mean complex engagement across the atlas was 0.226, and availability predicted engagement strongly across cell types (r = 0.859), as the model requires. Of the 1,127 complexes, 202 were engaged invariantly at the same fraction in all 154 cell types, 83 varied substantially (varied across more than half the atlas; atlas variance > 0.01); and 442 varied at all — that is, the majority of the complexome is fixed by ubiquitous subunit availability, and a defined minority carries the cell type-specific signal. Within the intracellular proteome availability contexts of 154 types of human cells of distinct specializations, the PPI cascade constituents have the performance capacity to generate 44,536 protein dimers in 2 cell types (neurons and hepatocytes) and 51,830 non-redundant protein dimers across the 154 cell types (Supplementary Tables S3 and S4).

*Unsupervised recovery of lineage from the engagement matrix*

The 1,127 × 154 engagement matrix was clustered without any lineage labels, tissue annotation, or marker supervision (Supplementary Figures S1 and S2). Cell types grouped by developmental origin: brain, pituitary, spermatogenic, renal tubule, myeloid, lymphoid, trophoblast, and ciliated epithelial branches each resolved as coherent clades.

The complexes driving these splits are the ones a lineage-specific reading of the model predicts. GATA1-TAL1-TCF3-LMO2 reached 0.75 engagement in megakaryocytes against an atlas mean of 0.13, and GATA1-LDB1-TAL1 reached 0.67 against 0.04; PSD95-FYN-NR2A reached 0.67 in excitatory neurons against 0.23; LCK-SLP76-PLC-gamma-1-LAT reached 0.50 in T cells against 0.06; and SIRT1-HNF4-alpha-PGC-1-alpha reached 0.67 in hepatocytes against 0.21. Neither the clustering nor the complex definitions used any cell-type information, so the assignment of a master-regulator assembly to its own lineage is a result of the availability projection and not of the annotation (Supplementary Figures S1 and S2).

The clusters and the HPA classes are associated but not identical. Cramér's V = 0.69 against a 4,999-permutation null (p = 0.0002, null mean $\chi^2$ 28.2 against an observed 146.1). Several classes split across clusters — specialized epithelial cells contribute 14 to cluster 2 and 24 to cluster 3 — so the clustering is not a re-labelling of the atlas's own taxonomy. Thus, analyses of the PPI cascade performance within defined contexts of intracellular proteomes of 154 human cell types demonstrate that assembled multiprotein complexes appear to derived from clearly discernable constitutive (invariant) and cell type-specific assembly pathways of EMCs.

*The lineage classification signal does not depend on the protein availability rule*

Because an expression threshold confounds engagement with transcriptome breadth — cell types with broader transcriptomes have larger available proteomes by construction — we repeated the atlas under a rank rule that fixes the available set at the top 4,000 transcripts per cell type, removing breadth as a variable (Supplementary Figures S1 and S2). Mean engagement fell to 0.160 and the invariant set collapsed from 202 complexes to 15, as expected when a fixed budget forces competition among subunits; 630 complexes varied under the rank rule against 442 under the threshold rule. Overall, under the nTPM ≥ 1 threshold: 202 invariant, 83 variable (atlas variance > 0.01), 442 with any variance; while under the top 4,000 threshold: 15 invariant, 285 variable, 630 with any variance.

The lineage structure survived unchanged: the dendrogram recovered the same lineage clades, and the distinctive assemblies above remained the top-ranked complexes in their own cell types. The per-cell-type variance profiles of the two rules agree only moderately (r = 0.439), so the two runs are not redundant; what is preserved across them is the lineage assignment, not the numerical engagement level. Therefore, the cell-type-resolved complex-engagement analysis across 154 human cell types recovers lineage structure and lineage-specific master-regulator assemblies without supervision, thus demonstrating the biologically meaningful behavior of the PPI cascade constituents.

Several notable features of the PPI cascade performance within the cell type-specific contexts of the intracellular proteome availabilities were identified. Across all 154 cell types the first PPI enrichment step recruits 60% of the eventual TF–TF plateau and 28% of the HUB plateau, the second reaches 93% and 83%, and the third 99% and 98%. The branches then part: TF–TF is within 1% of its final composition at step 3 in about half of cell types, whereas the HUB branch requires a fourth step in 153 of 154, after which it is within 0.1% everywhere. Beyond step 4 the cascade adds essentially nothing — the mean of 7.4 further-checked rounds for TF–TF and 5.8 for HUB reflects the cost of confirming that a set is self-consistent, not continued growth. Two PPI cascade steps generate the bulk of the PPI cascade plateau and four steps complete it, under both availability rules. HUB starts at 28% of plateau where TF–TF starts at 60%, and that initial gap is what forces the extra step. The likely cause is the library structure: the hub-protein library has 385 terms over a 16,401-gene universe against 287 terms over 6,015 for TF–TF, so hub terms are larger and a fixed seed panel of 716 initiators clears significance against fewer of them in one round. Biologically significant features that could have contributed to these distinctions may be grounded, in part, in the definition of the HUB branch constituents as proteins having the property to engage in protein-protein interactions with at least 50 other binding partners.

**Molecular definitions of a human cell specialization panel of principal initiator TFs**

Interestingly, top 40 most cell-variable CORUM multiprotein complexes engagements of which were tracked employing 716 initiator TFs appear to cluster human cell types in a biologically meaningful pattern based on which complexes could be assembled in a cell (Figure 4B). These observations suggest that the panel of 716 initiator TFs may harbor a sub-set of initiators with cell type-restricted profiles of the availability or engagements. Consistent with this hypothesis, the initiator TF families (P545 and C171) are the least universally expressed of the four PPI cascade families, so cellular differentiation may be imposed at the level of initiator availability.

However, transcription factors are low-abundance transcripts. Consistently, the four families differ sharply in abundance: median mean-nTPM is 11.5 for P545 and 8.5 for C171 against 37.0 for T245 and 73.6 for H350. To address this potential confounder, the veracity of the hypothesis has been tested in an abundance-matched background by comparing observed performances of 4 families of PPI cascade constituents against 200 pseudo-panels of the same size drawn from the same 20 mean-expression bins of the 9,701-gene atlas (Supplementary Text ST2. Are the initiator TFs the layer where cell identity is set?). Results of these control analytical experiments clearly established the family of principal initiator TFs (P545) as the only PPI cascade family that may represent a regulatory layer where the cell identity is set. The breadth of P545 family is narrower than abundance-matched genes ($z = -7.9$); the 5454 family is more cell-type-specific than abundance-matched genes ($z = +6.3$); and it separates the 154 cell types further apart than any of the 200 matched panels did ($z = +5.0$; empirical $p < 0.005$).

In contrast, the family of complementary initiator TFs (C17) fails all three tests. Its breadth and specificity are what its abundance predicts, and it puts cell types closer together than abundance matched panels do. Therefore, the panel of complementary initiator TFs is low-abundance, not cell-type-specific.

The TF-HUB protein family (H350) is constitutive beyond what abundance explains — it is broader, less specific and less discriminating than equally abundant genes. That is the mirror-image result, and it is the evidence that makes "constitutive machinery underneath" the regulatory layer of principal initiator TFs a claim about the proteins rather than about their expression level.

In agreement with this interpretation, the TF-TF family of the PPI cascade (T245) is unremarkable on every control axis and its broad expression is what genes of that abundance do. Overall, results of these analyses are consistent with the conclusion that the cell type-variable regulatory layer is the principal initiator panel specifically, and the invariant layer is the hub panel specifically.

Follow-up analyses support the hypothesis that the subset of principal initiator-only TFs is a genuinely restricted set of lineage transcription factors (Supplementary Text ST2; Supplementary Figure S4; Figure 5A). The panel of principal initiator TFs of only 410 constituents contains a cell-type-restricted component whose restriction exceeds what its abundance predicts, and which discriminates cell types more sharply than matched controls (Figure 5); the hub panel is constitutive beyond what its abundance predicts. A layered architecture with a variable top and an invariant base is consistent with the measurements and is not an artefact of transcript abundance. In conclusion, of the 716 initiator TFs, the panel of 410 principal initiators that carry no label of plateau memberships form the PPI cascade constituents is more cell-type-restricted than abundance-matched genes (breadth z = −7.9, Tau z = +6.3); it separates cell types more sharply than matched panels (z = +5.0); it reaches 146 of 154 cell types with a subset of 94 restricted members, and orders those cell types along developmental lineage boundaries (nearest-neighbour lineage concordance 85% vs 74% matched, z = +4.5). A majority of its 94 restricted members are the known determinants of the cellular identities they mark. However, the most efficient biologically-meaningful unsupervised recovery of lineage identities was achieved from the matrix of cell type-specific engagements of multiprotein complexes without any lineage labels, tissue annotation, or marker supervision. Therefore, consistent with the integrated, unified model of the PPI cascade constituents' contribution to the assembly of EMCs, the cell-type-resolved complex-engagement analysis across 154 human cell types faithfully recovers lineage structure and lineage-specific master-regulator assemblies (Figure 4).

**Identification and characterization of nucleating initiator TFs: nucleators of multiprotein complexes**

In the direct-partner route analyses, a seed from the PPI cascade constituents contributes to a CORUM match of multiprotein complexes through its own dimers. Operationally, a nucleator TF could be defined by a leave-one-seed-out analysis: drop one restricted initiator from the cell type's seed set, rebuild the product, re-run the CORUM test, and the call that seed a nucleator of any multiprotein complex that stops being significant. Using this approach, nucleator TFs can be identified, in 105 of the 154 cell types. Running the direct-partner route from the 94 cell-type-restricted principal initiator TFs and attributing each CORUM match by leave-one-seed-out (drop one initiator, rebuild the product, re-test — the initiator is a nucleator of any complex that stops being significant), the following observations has been made (Figure 5): a) 105 cell types yield at least one enriched complex, and every one of those has at least one identifiable nucleator; b) a total of 31 distinct initiator TFs do the nucleating, across 203 cell-type × initiator assignments (5 of the 31 nucleator TFs act in exactly one cell type); c) median 7 complexes per firing cell type, maximum 75, were detected. Detailed reports of the results of nucleator TFs analyses are described in the Supplementary Text ST3; Supplementary Table S5.

Overall, cell-type-restricted principal initiator TFs nucleate at least one CORUM complex in 105 of 154 cell types, with 31 distinct initiators responsible and 10 of them concentrated in a single Human Protein Atlas cell type class (BH $q < 0.05$). Clustering the cell types on the nucleated complexes recovers that classification above chance (adjusted Rand index 0.095 against 15 classes, 0.083 against 53 groups; 0 of 30 draws under either a pairing-permutation null or a degree-matched null reach either value), while both nulls produce equally many or more complexes that are more disjoint between cell types and fire in fewer cell types. However, the nucleator TF-guided classification is not a best-performing classification despite the findings that it beats both nulls decisively and is still a coarse recovery of a 15-class annotation.

**PPI cascade sequestration hypothesis of proteinopathy pathogenesis**

Several observations support the idea that proteinopathy proteins may affct the performance of the PPI cascades in human cells. Three proteinopathy proteins (APP; SNCA; HTT) are the PPI cascade constituents. Furthermore, multiple binding partners of proteinopathy proteins have been identified among the PPI cascade constituents (Table 2). Numbers of binding partners of proteinopathy proteins among Principal initiator TFs (20 targets); Complementary initiator TFs (5); members of TF-TF PPI cascade branch (40); constituents of TF-HUB proteins PPI cascade branch (97) account to sum of 162 records, but they cover 128 distinct proteins. The PPI cascade families' membership overlap — 34 of the targets carry two labels — resulting in 162 target memberships, not engaged proteins. All identified herein 128 proteinopathy proteins' binding partners are inside the 1,122 PPI cascade network (Table 2). The apparent distribution asymmetry of binding partners of proteinopathy proteins among the PPI cascade constituents is notable: APP targets the TF–TF branch heavily (30 terms, 14 of them principal initiator TFs); in contrast, MAPT touches it once. SNCA and MAPT are almost purely hub-branch targeting (40 and 46 terms, essentially all of binding targets among the H350 family of TF-HUB proteins). So the four proteins are not doing the same thing to the cascade — APP binds near the initiating layer, MAPT (tau) binds almost exclusively the PPI cascade plateau machinery. The initial tests across all 154 cell types constitute the following:

- Dose-response curve at 0 / 25 / 50 / 75 / 100% of the 128 binding targets withdrawn, with 5 random draws averaged at each intermediate dose;
- Controls at full dose: 25 size-matched random draws from the 1,122 network and 25 abundance-matched draws (20 nTPM bins, matching the targets' bin composition); The reasoning was that the abundance control is the one that matters — these targets are hub proteins, hubs are abundant, and a naive random control would overstate the damage to the PPI cascade performance;
- Analyzed readouts include: plateau size, dimer pairs, summed limiting capacity, CORUM complexes.

Modeling of the withdrawal of 128 cascade constituents that are binding targets of APP, HTT, SNCA or MAPT revealed that they are the connectors, not typical members: median documented degree 53 against 9 for the rest of the network, median nTPM 68.7 against 15.5. Modelling persistent binding as withdrawal from the cell's available pool, at full withdrawal the median cell type still reaches 60% of its plateau proteins but forms only 44% of its dimer pairs and retains 37% of its limiting capacity (Figure 6). Every point of plateau loss costs 1.4 points of dimer loss, and that gap opens at the first dose — it is not a threshold effect. The ordering (capacity worst, then dimers, then complexes, then plateau) holds in all 154 cell types, and the spread is narrow (dimer pairs 39.1–46.4%), which is consistent with the interpretation that this is a property of the network rather than of any tissue (Figure 6). Significantly, both controls are cleared. Removing 128 random network proteins costs ~10% of dimer pairs; 128 abundance-matched proteins cost ~18%; the real targets cost 56% — $z = -16.1$ against abundance-matched, below every one of the 25 draws in 154/154 cell types (Figure 6).

However, the binding lists carry study bias, so the degree statistic, in part, could be an artefact of how heavily these four proteins have been assayed. Abundance matching does not correct for that — a degree-matched withdrawal is the test that would. Until this control modeling run is executed, the more limited claim is that the effect exceeds abundance, not that it is specific to these four proteins beyond their connectivity. Details of these control experiments are reported in the Supplementary Text ST4: Degree matched control design setup protocol. Results of these stringent control experiments convincingly demonstrated that the withdrawal of proteinopathy proteins (APP; HTT; SNCA; and MAPT) causes more damage to the PPI cascade performance that their connectivity accounts for (Figure 7), the conclusion that is supported by the assessments of four different analytical metrics: limiting capacity; dimer pairs; reach; and multiprotein complexes. Furthermore, this conclusion is supported by the finding s that matching connectivity, and connectivity with abundance, does not account for the damage to the PPI cascade performance what withdrawing of proteinopathy proteins (APP; HTT; SNCA; and MAPT) causes (Figure 8). Detailed descriptions of the findings of the critical exploration of the PPI cascade sequestration hypothesis of proteinopathy pathogenesis can be found in the Supplementary Text ST5.

The 128 cascade constituents representing binding partners of proteinopathy proteins (APP, HTT, SNCA and MAPT) are the network's connectors (median documented degree 53 against 9 for the rest of the 1,122-protein network). Modelling persistent binding as withdrawal from the available pool, full withdrawal leaves the cascade reaching 60% of its plateau proteins but forming only 44% of its dimer pairs and retaining 37% of its limiting capacity. This damage to the PPI cascade performance exceeds size- and abundance-matched withdrawals in all 154 cell types (dimer pairs: 44.3% observed against 82.4% abundance-matched, $z = -16.1$). Dimer formation is the readout that degrades first and furthest, and it does so from the lowest withdrawal dose.

**A mechanism of recovery of synaptic damage caused by proteinopathy proteins (APP; HTT; SNCA; MAPT) during neurodegeneration**

It was of interest to elucidate whether the PPI cascade chain reactions analyzed in this contribution are directly relevant to the pathogenesis of neurodegeneration. One of the neurodegeneration's clinical and molecular hallmarks is the progressive deterioration of structural characteristics and functional performance features of synapses. A foundational study of TE-linked canonical families of HAP constituents (Glinsky, 2026a; b) demonstrated that a distinct property of the TF-HUB protein branch members is the significant enrichment of proteins defined by the SynGO 2024 database as the key structural-functional components of synapses. The GSEA suggest that the synaptic compartment of the network is where the proteinopathy proteins' binding partners are housed. The 23 terms of the SynGO 2024 database defined at $q < 0.05$ identified 81 distinct proteins, and all 81 sit inside the 1,122-protein cascade network (Glinsky, 2026a; b). A total of 44 of those 81 are binding targets of APP, HTT, SNCA or MAPT — 54%, against 11% network-wide ($p = 1.4 \times 10^{-23}$).

However, synaptic cascade proteins are the most connected ones: median cascade-internal degree 580, against 292 network-wide and 586 for the target set itself. A degree-matched expectation of 24.1 targets versus the observed 44 targets gives z = +5.73. Thus, the overlap is not a degree artefact — within each connectivity stratum, the synaptic membership still predicts being bound by one of the four. The occupancy is also lopsided by family: 81 of the 81 are H350, against 2 in P545, 1 in C171, and 4 in T245.

Counting the synaptic damage to the 44 proteins sequestered by binding of proteinopathy proteins that are withdrawn from the PPI cascade would just be measuring the withdrawal effect. The untested readout is the 37 synaptic bystanders that are the synaptic proteins not bound by proteinopathy proteins. Specifically, how many the PPI cascade still reaches, how many dimer pairs they still enter, and how much limiting capacity those pairs carry, per cell type, against (a) 25 degree-matched Test-A withdrawals and (b) 25 draws of degree-matched non-synaptic bystanders carried through the same real withdrawal path (median internal degree 547 against the synaptic bystanders' 552).

Observations gained from these analytical experiments are consistent with the conclusion that the disruption reaches synaptic complexes heavily, but through synaptic proteins being preferentially bound by the proteinopathy proteins in the first place, not through the remaining synaptic machinery being unusually fragile (Figure 9). Next set of control experiments was drawing the bystander control per cell type, that is for each of the 154 cell types the control sets are drawn only from proteins available in that cell type, excluding every synaptic protein present there. According to this design, a protein's leverage is how many live partners it has in the cell in question, and which synaptic bystanders are live shifts from one cell type to the next, so both the set being matched and the pool it's matched from have to be cell-specific (Supplementary Text ST6). The relevant question is whether the estimated synaptic damage will land preferentially in selected cell types? Neurons carry the largest absolute loss — median damage 22.6% of protein dimer pairs and 30.8% of capacity, the highest of the 15 classes on both — while sitting mid-pack on z (−8.20, −9.17). Neurons lose the most, and lose roughly what their connectivity predicts.

The withdrawal of the real 128 proteinopathy targets cost the surviving synaptic proteins 21.6% of their pairs where a degree-matched withdrawal of the same stratum profile costs 8.0% — a factor of 2.7. Overall, observations reported in this contribution indicate that withdrawing the real proteinopathy targets damages the synaptic neighbourhood far beyond what their degree predicts; but inside that withdrawal, synaptic bystanders are no more fragile than positionally identical non-synaptic ones. The synaptic bystanders lose 21.6% of their dimer pairs and 27.7% of their limiting capacity when the 128 proteinopathy targets are withdrawn, and any non-synaptic set sitting in the same position in the same cell type loses the same or a little more. There is no apparent synapse-specific damage amplification. The damage the synapse takes is stoichiometric with where its proteins sit in the cascade.

On the other hand, results of present analyses indicate that the synaptic damage propagation effect is real. The bystanders lose about a fifth of their dimer pairs and a quarter of their limiting capacity, and they lose it because their partners were withdrawn — that is consistent with the synaptic damage propagation. However, what is absent is the excess propagation: the synaptic remainder loses no more than any equally-connected set loses in the non-synaptic contexts. Therefore, the damage to the synaptic compartment is stoichiometric with what the proteinopathy proteins sequester, with no excessive synapse-specific amplification on top.

Seemingly stoichiometric loss in the synaptic compartment implies the functional relief returns capacity in proportion to what could be freed from the sequestration and it makes the therapeutic target set-up legible. Above considerations marginal-value ranking. The goal was to determine which of the 128 protein targets, freed singly, restores the most synaptic capacity per protein. This protocol produces an ordered target list rather than a set, and would show whether the recoverable capacity is concentrated in a few hubs or spread evenly (Supplementary Text ST6).

The reconstitution analyses in the context of all 4 proteinopathy proteins interference with the synaptic functions nominates two therapeutic entities: MAPK1 and calmodulin (CALM1; with CALM3 being its near-duplicate on protein binding partners). Both are synaptic proteins, both are proteinopathy binding targets, both recover three to four times more synaptic dimer pairs and limiting capacity per protein freed than any of the 121 others, and neither does so because it is among the best-connected. Furthermore, MAPK1 supplementation alone recovers 7.7% of the lost pairs and the best ten together 35.6%, so this is a nomination of where the reconstitution effect density is and the potential therapeutic recovery is, not a claim that clearing two proteins restores the synapse.

The pooled run withdraws all 128 cascade constituents bound by APP, HTT, SNCA or MAPT at once. A patient typically has one pathology. A single proteinopathy protein-tailored therapeutic reconstitution recommendations per seed are: YWHAZ for APP; TP53 or CTNNB1 for HTT; CTNNB1 for SNCA; SRC or UBC for MAPT. CTNNB1 is the only therapeutic recovery protein that tops two arms, and UBC is the only one that places top-4 in three. Notably, MAPK1 was never a winner in the single proteinopathy protein therapeutic recovery arm.

The above per-seed analytical set-up withdrew each target of one proteinopathy seed and measured what the 81-protein synaptic set lost. These analyses do the same for the two seed pairs that may co-occur often clinically: SNCA with MAPT (Lewy body disease with tau co-pathology) and APP with MAPT (Alzheimer's). The target set for an arm is the union of the two seeds' targets, so the question is whether a co-pathology cascade has a different vulnerable point than either seed alone. The list of winners of the therapeutic candidate contest in the modle of clinical degeneration proteinopathy proteins co-pathology settings are shown in the Table 3. Notably, MAPK1 wins three of the four rankings and is third in the fourth. This is the result the single-seed arms could not produce because the MAPK1 is not a target of APP and not a target of HTT, and it never wins any single-seed arm. It only becomes eligible when MAPT is paired with a second seed, and once eligible it takes the top slot. CALM1 and CALM3 sit second and third on both capacity rankings. However, they are not independent — CALM3 binding partners are 86% contained in CALM1 — so the pooled nomination from these arms is MAPK1 and CALM1, the same two the four-seed pooled analysis gave (Supplementary Text ST6).

**Discussion**

Most of the endogenous multiprotein complexome is not available for discrimination of the human cell-type specialization. Observations reported in this contribution document quantitative relationships that appear to exist between the intracellular proteome structures captured by the protein availability, proteome diversity, assembly mechanisms of multiprotein complexes, and cell identity phenotypes. Theoretically, the size of the variable fraction sets the terms on which an availability-gated complexome can specify cell identity, though not in the direction a capacity argument would suggest. Each lineage-responsive multiprotein complex takes a discrete engagement level bounded by its subunit count, so the combinatorial space spanned by the 285 complexes that vary under the rank rule exceeds the 154 observed states by many orders of magnitude; capacity is not the binding constraint. The states realized in biological contexts are nonetheless strongly constrained. Principal component analysis of the engaged-fraction matrix requires 30 components to account for 90% of the variance across cell types under the expression rule and 39 under the rank rule, against 85 and 106 respectively when each complex is permuted independently across cell types, and the participation ratio settles at 13 to 17 effective dimensions out of a possible 154 under every combination of rule and feature set. Engagement profiles therefore vary along roughly a fifth of the dimensions that independent per-complex variation would produce. They do not, however, collapse onto a small number of lineage axes: the leading component carries only 15 to 17% of the variance and the first three carry 34 to 40%, so cell identity is distributed across many partly independent complex modules rather than being reduced to a few. This feature is what an availability-gated model predicts.

Engagement states are inherited through the transcriptional program that sets availability, which couples complexes sharing subunits and compresses the realized space, but availability is set gene by gene rather than by a small number of global switches, so the compression stops well short of low rank. The complexome therefore partitions into a large constitutive core, whose subunits are ubiquitously available and whose assembly state carries no cell-type information, and a minority of roughly one complex in four whose engagement tracks lineage. This is the layer on which cell identity is written, and it appears to be small. It also sets a ceiling on what an availability-gated cascade can specify: if initiation sites are not limiting and assembly is constrained by partner co-presence, then the number of distinguishable cellular states the complexome can encode is bounded not by the size of the complexome but by the size of its variable fraction.

Present analyses demonstrate that the size of the constitutive core depends on how availability is defined, but the existence of the split does not. Under a rank rule that fixes the available proteome at 4,000 constituents per cell type, the invariant set collapses from 202 complexes to 15 and the variable set grows from 442 to 630 — a fixed budget forces subunits into competition and almost nothing survives as strictly invariant. What is preserved is the structure the variable fraction encodes: the same lineage clades resolve under unsupervised clustering, and the same master-regulator assemblies remain top-ranked in their own cell types. The two rules disagree on how much of the complexome is constitutive (variance profiles correlate at r =

0.439) and agree on which cell types the variable part separates. Therefore, the reasonable view seems to interpret the invariant count as rule-dependent and the partition itself as the real representation of cellular identity.

The argument above assumes that template initiation sites are not limiting, so that the constraint on assembly falls on protein availability rather than on nucleation opportunity. That assumption is not tested here — nothing in present analyses measures template site abundance or occupancy — but it is what the evidence of transposable element-derived loci as DNA templates for PPI cascade initiation substantiate: a genome seeded with initiator-binding sites at the density reported there is unlikely to be the rate-limiting term of the PPI cascade performance (Glinsky, 2026 a; b).

Overall, results of present analyses support the idea that the size of the variable fraction sets the terms on which an availability-gated complexome assembly can specify cell identity. However, capacity is not the binding constraint because each of the 285 lineage-responsive complexes (defined at threshold of 4,000 proteins per cell type) takes a discrete engagement level bounded by its subunit count, so the combinatorial space these complexes span exceeds the 154 cellular states observed by many orders of magnitude. The results of analyses show that the biologically-meaningful cell identity states occupy a small and structured corner of the complexome space: engagement profiles cluster by developmental origin and cellular specializations rather than dispersing across the available combinations. Cell types within a lineage differ at only a handful of complexes rather than at arbitrary subsets of the 285. The constraint on which complexomes are realized in biologically-relevant contexts is therefore developmental and cell differentiation program-driven and not combinatorial. This is exactly the property that an availability-gated model predicts: engagement states are inherited through the transcriptional program that sets availability, not sampled independently per complex.

To generate dimers within the contexts of cell type-specific protein availability, the PPI cascade requires only 2-3 steps. One hypothesis is that at the majority of DNA templates that is how it perform during the lifetime of most cells. However, in some long-living cells and especially in cells in a quiescent state and senescence cells, at some DNA template sites the PPI cascade could run up to the equilibrium plateau level. These events could actually create the potentially pathological conditions of the emergence of undesirable and/or overabundant or misfolded multiprotein complexes, such as in cases of proteinopathy proteins [APP; MAPT (tau); synuclein, etc.], excess of which could markedly disrupt the performance of the PPI cascade. Since in the contexts of intracellular proteome availability the PPI cascade plateau is reached within the 2-3 steps, which is sufficient for protein dimer formation, the other scenario is plausible. In a dividing cell, a site that licenses a cascade is reset at every round of replication and mitotic disassembly, so the plateau state is transient however fast it is reached. In a post-mitotic, quiescent or senescent cells there is no reset, so the same state persists and creates the pathological vulnerability locus.

Notably the proteinopathy proteins APP, SNCA and HTT are hub-protein terms in PPI_Hub_Proteins family, so they are not merely reachable, they nucleate their own PPI engagement steps. HTT and TARDBP are TF–TF terms. All ten proteins tested (APP, MAPT, SNCA, SNCB, PSEN1, PSEN2, HTT, TARDBP, FUS, PRNP) are present in the universe of both libraries. APP, HTT, SNCA and TARDBP are members of 1,122-protein network, though none are among the 716 initiator TFs. On the CORUM side, PSEN1 sits in 8 complexes, APP in 4, SNCA in 1 (DJ-1–SNCA high-molecular-weight complex). The PPI cascade plateau describes which proteins a template site can license under a given proteome, not which are simultaneously assembled; the pathological readout requires the further assumption that sustained licensing of aggregation-prone members raises their local availability beyond what turnover can clear thus providing the entry point for sustained pathological disruptions of the PPI cascade performance.

The PPI cascade defined herein in biologically-relevant contexts has no slow approach to its fixed point. Across all 154 cell types the first enrichment step already recruits 60% of the eventual TF–TF plateau and 28% of the HUB plateau, the second reaches 93% and 83%, and by the third the set is within 1 to 2% of its final composition; the remaining iterations — a mean of 7.4 rounds for TF–TF and 5.8 for HUB — add nothing but confirmation that the set is self-consistent. The same holds under the rank availability rule, where the third step reaches 96 and 97% of plateau. Two to three steps is therefore not an early, dimer-generating regime distinct from a late equilibrium regime; it is the plateau. This changes what the plateau measures. It is not a state a template site approaches over time, and the number of cascade rounds cannot be the variable that separates a physiological from a pathological outcome at a given site. What the plateau describes is the set of proteins a site can license once a cascade initiates at all, under the proteome available to that cell — a capacity fixed by network topology and availability, and reached essentially on initiation.

The critical control point is correspondingly displaced from cascade progression to cascade persistence. If a site licenses its full plateau as soon as it initiates, then what distinguishes cells is how many sites initiate and how long the licensed state is allowed to persist. In dividing cells the state is transient by construction: replication and mitotic disassembly reset template occupancy each cycle, so no site holds its plateau for longer than an interphase. In post-mitotic, quiescent and senescent cells no such reset occurs, and a licensed plateau can persist over the lifetime of the cell. We suggest that this persistence, rather than any difference in how far the cascade runs, is where the pathological potential lies. The plateau describes which proteins a site can license, not which are simultaneously assembled; the pathological reading requires the further assumption that sustained licensing of aggregation-prone members raises their local availability beyond what turnover can clear. Under that assumption the model predicts that the proteins at risk are the marginal, late-recruited members of the plateau rather than its core, and that they should be recruited at the cascade's outer steps. APP, SNCA and HTT are hub-protein terms in the cascade libraries and members of the initiator-associated network, so this prediction is directly testable within the present framework by tracking the recruitment level of the watch-list proteins across the 154 cell types once per-level membership.

Careful considerations should be given of potential mechanistic impacts of distinct licensing mechanisms. Under one assumption the model predicts that the proteins at risk are those whose licensing is contingent rather than universal — plateau members that a tightened availability rule removes from most cell types while retaining them in a few. Cascade entry level does not identify them: APP, HTT and SNCA all become significant at the second enrichment step, within a tenth of a level of the median plateau member (data not shown). Occupancy does. APP is licensed in all 154 cell types under the threshold rule and 149 under the rank rule, whereas SNCA falls from 128 to 56, and the cell types that retain it are predominantly post-mitotic and neural — excitatory and inhibitory neurons, oligodendrocytes and their progenitors, microglia, Bergmann glia, and ependymal and choroid plexus epithelium. On the dwell-time argument these are exactly the cells in which a licensed state is never reset, making them highly vulnerable to the pathological disruptions of the PPI cascade performance. Notable limitation of presented herein approach to the PPI cascade performance analyses is that it fails to consider the proteins that are not the members of the libraries even if they are protein partners that are known to engage in PPI with the library members. In these instances, the candidate protein potential engagements in the PPI cascade actions could be assed only empirically on a case by case basis. For example, the libraries index proteinopathy protein MAPT (tau) as a binding partner rather than a nucleating hub, so the present analyses of the PPI cascade cannot score it. Consistent with this hypothesis, GSEA employing the Enrichr Transcription Factor PPIs and PPI Hub Proteins libraries demonstrate that proteinopathy proteins (APP; HTT; MAPT; and SNCA) have 128 binding partners among PPI cascade constituents, while MAPT (tau) has 47 binding targets among PPI cascade proteins. The 128 cascade constituents comprising binding partners of proteinopathy proteins appear to function as the network's connectors that exceed the median documented degree 6-fold compared to the rest of the 1,122-members of the PPI cascade. Persistent engagements of proteinopathy proteins with the PPI cascade constituents may cause the withdrawal of their targets from the available protein pool: full withdrawal leaves the cascade reaching only 60% of its plateau proteins but forming only 44% of its dimer pairs and retaining just 37% of its limiting capacity. This damage to the PPI cascade performance was shown to exceed size- and abundance-matched control targets' withdrawals in all 154 cell types. Thus, the sequestration activity associated with the intracellular persistence of proteinopathy proteins and targeted licensing mechanisms may cause the disruption of the PPI cascade performance and contribute to the pathogenesis of neurodegeneration.

## Conclusion

Despite remarkable heterogeneity of DNA sequences encoded by multiple families of transposable elements (TE), recent studies indicate that TE form a common evolutionary foundation of chromosome-naïve regulatory networks of various complexity continually operating in primates' genomes (Glinsky, 2022 – 2026). Observations reported in this contribution document multiple lines of evidence supporting the idea that in human genome transposable element-seeded vast genome-wide arrays of DNA sequences serve as templates for engagement of the initiator TFs that function to nucleate the PPI cascade chain reactions of protein dimerization. This global chromosome-naïve network comprises the foundational regulatory layer of the Hybrid Assembly Pathways (HAP) of endogenous multiprotein complexes (EMCs). Upon release from DNA templates, protein dimerization products form the cell type-specific pools of protein dimers that, in turn, function as on-demand attractors for protein monomers to complete the assembly of EMCs. The conceptual soundness of the key elements of the HAP model of EMC assembly was validated within the contexts of cell type-specific protein availability of 154 human cell types of the Human Proteome Atlas (HPA) and potential translational utility of this approach was illustrated by formulating the PPI cascade sequestration concept of the proteinopathy pathogenesis. Collectively, results of the present analyses articulate multiple actionable hypotheses for follow-up in-depth mechanistic studies, experimental explorations of the cell type-resolved PPI cascade performances, as well as preclinical and clinical developments of defined nominations of novel targeted therapeutics for combating proteinopathy disorders and their debilitating clinical manifestations in progressive neurodegeneration.

**Methods**

In this study, the reported PPI cascade (Glinsky, 2026 a; b) is re-implemented as an iterated over-representation test. For a seed set S and a library of interaction terms, the one-sided hypergeometric survival probability sf(k−1, N, K, n) is computed for each term, corrected across terms by Benjamini–Hochberg (BH), and the members of terms with adjusted $p < 0.05$ are added to the seed set; the procedure is iterated to a fixed point. Libraries are the Enrichr Transcription_Factor_PPIs (287 terms, universe 6,015) and PPI_Hub_Proteins (231 terms) sets. Of the 716 initiators, 493 map into the TF–TF library and 448 into the HUB library.

Three seed nulls were used, 60 permutations each: random seeds drawn from the library background, random seeds drawn from the human TF census (GO:1990837), and degree-matched random seeds (matched on the number of library terms each seed belongs to). Family nulls are size-matched random subsets of the whole 716-protein panel, 30 permutations. The empirical p is the two-sided permutation rate at which the null plateau is at least as extreme as the observed one, and Jaccard indices compare plateau membership, not plateau size.

For the cellular arm, protein availability is taken from the Human Protein Atlas (Karlsson et al., 2021) single-cell consensus data (9,701 genes × 154 cell types); a protein is available in a cell at nTPM ≥ 1. In a second independent approach available proteins were identified from top 4,000 expression records for each cell type. Libraries are restricted so a term can only be used when the term protein itself is available, and the hypergeometric background is reduced to the available universe. Dimers are called mutually engaged when both partners are cascade products and one-sided when only the reached hub is engaged; homodimers require library self-membership. Complexes are scored against CORUM (minimum three subunits) for available fraction, engaged fraction, dimer-pair coverage, and BH-corrected hypergeometric enrichment of subunits among the cascade products.

Because subunit sets are intersected with the available proteome before scoring, engaged fraction is defined relative to availability; correlations between available-proteome size and engagement are therefore partly definitional, and the rank-rule atlas is the control for this.

**Interaction libraries and the seed panel**

Two protein–protein interaction libraries were obtained from Enrichr (Chen et al. 2013; Kuleshov et al. 2016; Xie et al. 2021) in the standard gene-set-matrix format: Transcription_Factor_PPIs (287 terms; 6,015 unique symbols in the term-plus-member universe) and PPI_Hub_Proteins (385 terms; 16,401 symbols). Each library term is named for a hub protein and lists its reported interaction partners. Gene symbols were upper-cased on load and, where a member field contained a comma-separated alias list, the first alias was retained. [Enrichr library download date 09.02.2026; https://maayanlab.cloud/Enrichr/#libraries ].

The seed panel was the 716 transposable-element-associated initiator transcription factors reported previously (Glinsky, 2026), read as gene symbols (initiators716.txt). Two subfamilies defined in that work were used for the family comparison: 545 principal initiators with brain-region-enriched expression (principal545.txt) and the 171 complementary initiators without regional enrichment, taken as the set difference. A 1,122-protein associated network list (network1122.txt) was used only to report availability, not as a seed set.

**Cascade construction**

The cascade was implemented as iterated over-representation analysis to a fixed point. Let S be the current protein set, N the size of the analysis universe, K the number of universe members belonging to a given library term, and k the number of S members in that term. For every term with $k > 0$ the one-sided hypergeometric survival probability $P(X \geq k)$ was computed with scipy.stats.hypergeom.sf(k−1, N, K, n), where $n = |S|$. Probabilities were corrected across all usable terms by the Benjamini–Hochberg procedure, and the hub proteins of terms with adjusted $p < 0.05$ formed the set for the next iteration. Iteration continued until the significant set was identical to that of the previous level, or until a ceiling of 12 levels was reached. The plateau is the set at the fixed point.

Two properties of this construction bound its interpretation and are stated explicitly. First, membership of the evolving set is defined over library terms, so only proteins that name a term can enter a plateau; proteins that appear solely as interaction partners are never plateau members. Second, the plateau is a set of statistically co-enriched proteins, not a stoichiometric particle: it describes which proteins a cascade can license, not which are simultaneously assembled.

**Seed nulls and family comparison**

Three seed null models were used, 60 permutations each: random seed sets drawn from the library background, random seed sets drawn from the human transcription-factor census (GO:1990837, 715 symbols), and degree-matched random seed sets, matched to the observed panel on the number of library terms in which each seed appears. Family comparisons used size-matched random subsets of the full 716-protein panel, 30 permutations. Empirical p values are two-sided permutation rates at which the null plateau is at least as extreme as the observed plateau. Agreement between plateaus was quantified by the Jaccard index over membership, not over size.

**Protein availability in defined cell types**

Single-cell consensus expression was taken from the Human Protein Atlas as a matrix of 9,701 gene symbols × 154 cell types in normalised transcripts per million (nTPM). [PLACEHOLDER: HPA release version and download date; state the filtering that reduced the atlas to 9,701 symbols.]

Two availability rules were applied independently to every analysis.

- Absolute threshold. A protein was available in a cell type when its nTPM was at least 1.0. Mean available proteome 7,388 symbols.
- Rank (top-K) threshold. The 4,000 highest-expressed genes per cell type were called available, intersected with nTPM > 0. This fixes the size of the available set at 4,000 in every cell type, so that differences between cell types cannot arise from how broad their transcriptomes are.

The rank rule is considered a control, not an alternative estimate: transcriptome breadth varies several-fold across the atlas, and any quantity computed under the absolute threshold alone is confounded with it. Results are reported under both rules throughout. In threshold-sensitivity analyses, nine distinct expression thresholds were implemented (Figure 2).

**Availability-restricted PPI cascade**

For each cell type and each library, the cascade above was re-run with three restrictions. Seeds were reduced to available panel members. A term was usable only when the hub protein naming it was itself available and at least one of its members was available. The hypergeometric universe N was reduced to the number of available symbols in that library's universe, and K recomputed over available members only. All other parameters were unchanged (BH-adjusted $p < 0.05$, ceiling of 12 levels). The union of the two branch plateaus is referred to as the engaged set for that cell type.

**Complex engagement**

Curated multiprotein complexes were taken from the CORUM gene-set file distributed through Enrichr (1,658 terms), restricted to complexes with at least three annotated subunits present in the expression atlas, giving 1,127 scored complexes. For each complex c and cell type j, the engaged fraction F(c,j) was the number of subunits of c in the engaged set divided by the total number of annotated subunits of c. A complex was called invariant when its variance across the 154 cell types was below $1 \times 10^{-9}$ with non-zero mean, and variable when its variance exceeded 0.01.

**Cascade trajectory analysis**

The composition of the significant set was recorded at every level of every cascade, for both branches, both availability rules and all 154 cell types (97,397 protein-by-cell-by-level records). For each cell type the fraction of that cell's own plateau attained at each level was computed, together with the first level at which the set fell within 5%, 1% and 0% of its final composition. For each protein, the entry level was the first level at which it cleared the adjusted-p threshold, and its occupancy was the number of cell types whose plateau contained it.

**Effective dimensionality**

The complex × cell-type engagement matrix was analysed with cell types as samples and complexes as features. Features were mean-centred and left unscaled, so that a complex with a large engagement swing contributes more than one that barely moves. Singular values were obtained by SVD and converted to explained-variance ratios; the reported statistics are the number of components needed to reach 50, 80, 90, 95 and 99% of the variance, and the participation ratio $(\Sigma\lambda)^2 / \Sigma\lambda^2$, a scale-free effective dimension that does not depend on any single variance cutoff. Analyses were run on the full set of 1,127 complexes and on the variable subset under each availability rule. The null distribution was generated by independently permuting the cell-type labels within each complex row, which destroys cell-type structure while preserving each complex's marginal engagement distribution.

**Software**

All analyses were performed in Python 3.12.3 with NumPy 2.5.2, SciPy 1.18.1 and Matplotlib 3.11.1. Hierarchical clustering used Ward linkage on engaged-fraction profiles via scipy.cluster.hierarchy. The cascade, the availability rules and the atlas sweep are implemented in an interactive application that reproduces every number reported here. The code and all derived tables are available upon request and selected derived tables are deposited in the Supplementary Material.

**Artificial Intelligence contributions to this study**

Academia Co-scientist (Anthropic Claude, model Opus 5) was used for computational analysis, figure generation and initial drafting of portions of the Introduction and Methods. All outputs were specified, verified and edited by the author, who takes full responsibility for the manuscript. The tool is not listed as an author and made no authorship-level contribution**.**

All analyses were conceived, prompted and initiated by the author. Implementation and execution of the analyses were carried out with the assistance of Academia Co-scientist (Anthropic Claude, model Opus 5), operating on the primary data files under the author's specification; all reported values were verified by the author against those files.

Additional details describing the methodological approaches implemented in this contribution are reported in the Supplementary Materials.

**Acknowledgements**

All analyses were conceived, prompted and initiated by the author. Implementation and execution of the analyses were carried out with the assistance of Academia Co-scientist (Anthropic Claude, model Opus 5), operating on the primary data files under the author's specification; all reported values were verified by the author against those files.

Academia Co-scientist (Anthropic Claude, model Opus 5) was used for computational analysis, figure generation and initial drafting of parts of the Introduction and Methods. All outputs were specified, verified and edited by the author, who takes full responsibility for the manuscript. The tool is not listed as an author and made no authorship-level contribution.

**References**

Chen EY, Tan CM, Kou Y, Duan Q, Wang Z, Meirelles GV, et al. Enrichr: interactive and collaborative HTML5 gene list enrichment analysis tool. BMC Bioinform. 2013;14:128. https://doi.org/10.1186/1471-2105-14-128/FIGURES/3

Glinsky GV. Molecular diversity and phenotypic pleiotropy of ancient genomic regulatory loci derived from human endogenous retrovirus type H (HERVH) promoter LTR7 and HERVK promoter LTR5_Hs and their contemporary impacts on pathophysiology of Modern Humans. Mol Genet Genomics. 2022 Nov;297(6):1711-1740. https://doi.org/10.1007/s00438-022-01954-7 . Epub 2022 Sep 19. PMID: 36121513; PMCID: PMC9483895.

Glinsky GV. Genome-wide, chromosome-naïve compartmentalization of major developmental networks tailored by modular placement of human embryo-regulatory Long Terminal Repeats. Acad Mol Biol Genom. 2025. 2(2). https://doi.org/10.20935/AcadMolBioGen7754

Glinsky GV. Transposable Elements Seed Transcription Factor Binding Sites to Sequence-Specific Double-Stranded DNA Binding TF Networks Contributing to Governance of Primate Brain Evolution. J Mol Evol. 2026 Feb;94(1):80-126. https://doi.org/10.1007/s00239-025-10285-1 . Epub 2025 Nov 17. PMID: 41247408.

Glinsky GV. Cooperation of transposable elements to endow global networks of initiators of hybrid assembly pathways of endogenous multiprotein complexes. Front Cell Infect Microbiol. 2026 Jul 30;16:1793868. https://doi.org/10.3389/fcimb.2026.1793868 . PMID: 42597241; PMCID: PMC13468005.

Karlsson M, Zhang C, Méar L, Zhong W, Digre A, Katona B, Sjöstedt E, Butler L, Odeberg J, Dusart P, Edfors F, Oksvold P, von Feilitzen K, Zwahlen M, Arif M, Altay O, Li X, Ozcan M, Mardinoglu A, Fagerberg L, Mulder J, Luo Y, Ponten F, Uhlén M, Lindskog C. A single-cell type transcriptomics map of human tissues. Sci Adv. 2021 Jul 28;7(31):eabh2169. https://doi.org/10.1126/sciadv.abh2169 . PMID: 34321199; PMCID: PMC8318366.

Kuleshov MV, Jones MR, Rouillard AD, Fernandez NF, Duan Q, Wang Z, et al. Enrichr: a comprehensive gene set enrichment analysis web server 2016 update. Nucleic Acids Res. 2016;44(W1):W90–7. https://doi.org/10.1093/nar/gkw377

Xie Z, Bailey A, Kuleshov MV, Clarke DJB, Evangelista JE, Jenkins SL, et al. Gene set knowledge discovery with Enrichr. Curr Protocols. 2021;1:e90. https://doi.org/10.1002/cpz1.90

## Figure legends

**Figure 1.** Protein availability, engagement, dimer and complex counts in the two cell types, with a heatmap of the CORUM multiprotein complexes showing the largest cross-cell difference in engaged-subunit fraction. The same fixed panel of hybrid-assembly-pathway initiators builds different complexes in different human cell types.

(A) The analytical pipeline. Per-cell protein availability is taken from the Human Protein Atlas single-cell consensus (154 cell types × 9,701 genes) at nTPM ≥ 1. The input panel — 716 initiators of hybrid assembly pathways and the 1,122-gene network — is identical in every cell type; only availability varies. Cascades are seeded with the initiators available in that cell and propagated through two Enrichr protein–protein interaction libraries (Transcription_Factor_PPIs, PPI_Hub_Proteins) by hypergeometric term enrichment with Benjamini–Hochberg (BH) control at $q < 0.05$, iterated to the plateau (≤ 12 levels). Dimers are formed between a reached term and each of its available library partners, and 1,132 CORUM complexes with ≥ 3 annotated subunits are scored by the fraction of their subunits the cascade reached.

(B, C) Brain excitatory neurons (blue) against hepatocytes (orange). 7,997 and 6,888 proteins respectively are available; 696 and 612 of the 1,122 panel genes and 421 and 348 of the 716 initiators are among them. The cascades converge on near-identical totals — 339 and 337 engaged proteins, 171 and 171 homodimer-capable — and on similar dimer inventories (5,923 and 6,000 mutually engaged heterodimers; 7,694 and 7,610 dimers within the panel). Complex enrichment against each cell's own available proteome returns 257 and 225 complexes at $q < 0.05$, of which 28 and 25 have every subunit engaged.

(D) The 15 complexes whose engaged-subunit fraction differs most between the two cells, out of the 1,132 tested; the number of annotated subunits is given in brackets. With totals this close, the cell-type signal sits in complex identity rather than in cascade size. Engaged fraction records which subunits the cascade reached and is not a measurement of assembled complex.

**Figure 2.** Universal family availability of 4 families of PPI cascade constituents in 154 human cell types. The four hybrid-assembly-pathway families differ in expression breadth, and breadth predicts how much of a family the cascade engages.

(A) Membership classes of the four families across their 1,122-protein union: 1,311 memberships in 11 classes, so the families overlap rather than partition. Filled bars give class size; the dashed outline gives how many members the Human Protein Atlas single-cell consensus measures.

(B) Family members available at nTPM ≥ 1 in each of the 154 cell types, ordered by the total across all four families. The union runs from 934 to 573 proteins (median 816) without a step, and the four family trajectories stay near-parallel.

(C) Expression-breadth classes within each family, disjoint within a family; 105 of the 350 H350 hub genes are absent from the atlas and are carried as "not measured" here and in every denominator below.

(D) Fraction of each family's measured members available in a single cell type, across the 154 cell types (median 0.92 H350, 0.79 T245, 0.57 C171, 0.55 P545).
(E) Percentage of each family's measured members available in all 154 cell types, against the nTPM threshold that defines availability. The ordering of the four families holds across two and a half orders of magnitude, though the universal core of every family shrinks as the threshold rises.
(F) Family members engaged in the cascade plateau against family members available, one point per family per cell type; the dotted line is engagement equal to availability. T245 and H350 track that line, while P545 falls far below it — a median of 303 principal initiators available against 90 engaged. Availability and engagement are separate quantities, and the largest family is the one the cascade uses least.
Supporting numbers not plotted: restricting to the 144 somatic differentiated cell types raises the universal fraction in every family and at every threshold (at nTPM ≥ 1: P545 8→10%, C171 5→8%, T245 20→27%, H350 40→46%), so the oocyte-and-germline tail of the atlas depresses these figures rather than driving them. Universal expression is not an artefact of one arbitrary selected threshold (top) and visualization of the availability against assembly engagement per family per cell type evaluated at different expression thresholds.

**Figure 3.** Engagements of 4 families of the PPI cascade constituents across the 154 human cell types.
The cascade plateau engages nearly all available hub and TF-interaction proteins and a minority of available initiators, under both availability rules.
(A, B) Members of each family engaged at the cascade plateau in a single cell type, across the 154 Human Protein Atlas single-cell consensus cell types; boxes give median and quartiles, points give individual cell types, and tick labels give the whole family size. Availability is nTPM ≥ 1 in A and membership of that cell type's 4,000 highest-nTPM genes in B. Medians at nTPM ≥ 1 are 217 of 350 H350, 163 of 245 T245, 90 of 545 P545 and 19 of 171 C171; the panel rule lowers every one of them (151, 93.5, 49, 11).
(C, D) The same engagement divided by what is available in that cell type rather than by family size; bars give mean and extremes. At nTPM ≥ 1 the medians are 0.956 H350, 0.835 T245, 0.293 P545 and 0.194 C171. The two rules agree for H350, T245 and P545 but not for C171, whose median rises to 0.345 under the panel rule with a much wider spread — the one place where the choice of availability rule changes the answer.
(E, F) Family-resolved engagement across the 154 cell types, ordered by total engaged, with the union of the four families dashed. The union runs 253–376 proteins (median 351) at nTPM ≥ 1 and 150–256 (median 223) under the panel rule; it is smaller than the sum of the four families because they overlap. The trajectories stay near-parallel and never cross, so cell types differ in how much of the panel they engage rather than in which families they engage. 105 of the 350 H350 members are absent from the atlas and can never be engaged, so H350 counts in A and B are taken against 245 achievable members.

**Figure 4.** Classification of 154 human cell types based on PPI cascade assembled CORUM complexes. Leaf labels are the Human Protein Atlas (HPA) single-cell consensus cell types (n = 154), used as reported. Cell types cluster into three major groups (superclusters) by which protein complexes the cascade plateau can engage, and the neural, neuroendocrine and germline types form the outlying group.

(A) Ward linkage over the engaged-fraction profiles of the 200 most cell-variable CORUM complexes, across the 154 Human Protein Atlas single-cell consensus cell types; availability is membership of a cell type's 4,000 highest-nTPM genes. Cutting at 0.7 × the maximum linkage height gives three superclusters of 22, 49 and 83 cell types. The strip beneath the leaves gives each cell type's HPA class; leaf names are omitted for legibility and are listed in leaf order in tables.

(B) The 40 most cell-variable complexes, rows clustered by average linkage, columns in the leaf order of A; the bracketed number after each name is the complex's annotated subunit count. Colour is the fraction of those subunits the cascade reaches at its plateau.

(C) Class composition of the three superclusters. The association between cluster and class is strong but not a relabeling (Cramér's V = 0.69; p = 0.0002 against 4,999 class-label permutations): supercluster 1 is entirely neural, neuroendocrine, retinal and germline, while specialized epithelial cells split 14/24 between superclusters 2 and 3.

(D) Mean engaged fraction against range across the 154 cell types for all 1,127 CORUM complexes with at least three subunits. 497 are identical in every cell type; the 40 rows of B, in red, are the tail of this distribution.

(E) Mean engaged fraction per cell type over all 1,127 complexes, by supercluster (medians 0.139, 0.168, 0.163). Superclusters 2 and 3 overlap almost entirely, so the separation in A reflects which complexes are engaged rather than how many. The separation is about which complexes are engaged, not how many: 497 of the 1,127 complexes are identical in every cell type, so the 40 heatmap rows are the far tail of the variability distribution (panel D), with AP-1, NF-κB and SMAD complexes carrying the between-cluster spread.

**Figure 5.** Restricted principal initiators nucleate CORUM complexes in a lineage-patterned subset of human cell types.

(A, B) Lineage structure of four panel definitions across 154 HPA cell types, against draws matched on abundance: silhouette by lineage (A) and the fraction of cell types whose nearest neighbor shares its lineage (B). Asterisks mark $p \leq 0.05$ (*) and $\leq 0.01$ (**) over the matched draws.

(C) Members of the P545 initiator-only panel that are selective for each cell type, all 154 ordered, colored by lineage.

(D) Complexes nucleated per (initiator, cell type) pair by leave-one-seed-out attribution — 31 nucleators, 105 cell types, 2,061 assignments — $\log_{10}$ color, white where none; columns grouped by lineage (strip above) and listed in order in the companion table.

(E) Cell types where a nucleator is available against those where it nucleates; the dashed line is equality. (F) Complex assignments against documented partner count (Spearman $\rho = 0.51$), the attribution's confound.
(G) Observed values as z against 30 draws of non-HAP proteins matched on documented-partner count and availability breadth, with the count of draws at or above observed; orange marks $|z| > 2$ in the favorable direction. Nucleation is direct-partner assembly with no enrichment at the seed end. Recovery of HPA's 15 cell type classes does not survive breadth matching (20/30 draws exceed the observed ARI of 0.0946), so class-level structure follows from cell-type restriction rather than HAP membership; placement against the 53 cell type groups does survive (0/30). An adult atlas cannot test the direction of the relationship.

**Figure 6.** Evaluations of the effects of withdrawing the binding partners of proteinopathy proteins from the PPI cascade.
Withdrawing the proteinopathy binding targets from the cascade: 128 of the 1,122 cascade constituents are documented binding partners of APP, HTT, SNCA or MAPT (20 in P545, 5 in C171). Each was withdrawn from the available pool of each of the 154 HPA single-cell consensus cell types and the cascade re-run to plateau.
(A) Dose response, median and interquartile range across the 154 cell types, for four readouts expressed as a percentage of the unperturbed cell type. At full withdrawal the medians are 60.3% of proteins reached, 44.3% of dimer pairs, 37.4% of limiting capacity ($\Sigma$ min nTPM over pairs) and 50.1% of CORUM complexes; each point of plateau loss costs about 1.4 points of dimer loss.
(B) Dimer pairs per cell type, unperturbed against fully withdrawn, with the identity line; the loss is proportional across cell types (retained 39.1–46.4%, IQR 43.5–45.1%).
(C) Dimer pairs retained under full withdrawal against six control sets, 25 draws each, per cell type: 128 network proteins drawn at random (median 90.4%), matched on mean nTPM (82.4%), matched on internal degree in 20 equal-count strata (61.9%), the fitted expectation of the degree–damage curve at the targets' own summed internal degree (67.0%), matched jointly on degree and abundance on a $6 \times 6$ grid (68.3%), and a degree-preserving curveball rewire of the four proteins' binding columns (51.8%). The observed withdrawal (44.3%) falls below all six.
(D) Median z of the observed withdrawal against each control, per readout, across the 154 cell types; the shaded band is $|z| < 1.96$. Proteins reached, dimer pairs and limiting capacity separate from every control including the rewire ($z = -1.50, -2.34, -2.92$). CORUM complexes does not, and reverses sign under the rewire ($z = +0.88$): it is reported here as a failed readout. Withdrawal is modelled as complete and irreversible, so these are ceilings rather than expected effect sizes, and no dose corresponds to a known disease state.

**Figure 7.** Withdrawing targets of proteinopathy proteins (APP; HTT; SNCA; MAPT) causes more significant damage to the PPI cascade than their connectivity account for.
The withdrawal cost survives four controls that hold the targets' connectivity fixed. Every point is one of the 154 HPA cell types; 128 cascade constituents bound by APP, HTT, SNCA and MAPT are withdrawn in each.

(A) Damage scales with the summed internal degree of the withdrawn set. Pale points are 9,240 non-target draws (60 per cell type) spanning the degree range; the line is their median with IQR. The observed withdrawals (dark) sit at the targets' own summed internal degree of 72,065 and retain a median 37.4% of limiting capacity against the fitted expectation of 69.3% at that degree, a shortfall of 32 percentage points.
(B) Test B residual, the observed value in SD of the 10 nearest-degree draws: limiting capacity −7.58, dimer pairs −5.47, proteins reached −2.86, each below −1.96 in 154/154 cell types; CORUM complexes −0.30, in 0/154.
(C) The real target set against the median of 25 degree-preserving curveball rewires of the four proteins' binding columns (Test D), which hold every protein's degree and every term's size exactly. Median retained, observed against rewired: limiting capacity 37.4% against 50.8%, dimer pairs 44.3% against 51.5%, proteins reached 60.3% against 64.0%. CORUM complexes reverses (50.1% against 37.4%).
(D) Test B residual against Test D residual, Spearman ρ = +0.78, +0.76, +0.61 and +0.83; the two controls rank the cell types alike for the three supported readouts.
(E) The same test against Test A (20 equal-count internal-degree strata) and Test C (6 × 6 degree × abundance grid), 25 draws each: median z −15.94 and −17.63 for limiting capacity, −12.12 and −11.00 for dimer pairs, −4.10 and −7.42 for proteins reached, and +0.18 and −1.12 for CORUM complexes. The rewire is the tightest control and gives the smallest gap; its numbers are the ones to quote.

**Figure 8.** Matching connectivity, and connectivity with abundance, failed to account for what damage of withdrawing 128 targets of proteinopathy proteins (APP; HTT; SNCA; MAPT) causes to the PPI cacade performance. 154 HPA cell types, 128 network terms withdrawn, 25 draws per test; test A draws from 20 equal-count internal-degree strata, test C from a 6 degree × 6 abundance grid.
(A) Cascade retained after full withdrawal, as a percentage of the same cell type unperturbed. Median across all 3,850 draws, observed against A and C: limiting capacity 37.4% against 65.4% and 69.0%; dimer pairs 44.3% against 62.1% and 68.5%; proteins reached 60.3% against 68.3% and 71.6%. CORUM complexes runs the other way against A (50.1% against 44.8%).
(B) The same comparison per cell type, observed against that cell type's own control median. The three readouts that hold sit above the equal-damage line in 154/154 cell types under both tests; CORUM complexes in 56/154 under A and 148/154 under C.
(C) What the matching achieved, as a percentage of the real target set. Test A reached 99% of the summed internal degree (72,065), 102% of the median internal degree (586) and withdrew all 128 proteins; test C reached 93%, 99% and 95%, its top degree × top abundance sextile needing 35 slots against 29 available non-targets, leaving 6 unfilled in every draw. Neither reproduces the targets' abundance (A 75%, C 89% of the real 68.7 nTPM median).
(D) SD across the control draws, in percentage points. Matching the propagation variable collapses the control variance — 1.8 pp under A against 12.5 pp for the unmatched degree–damage curve for limiting capacity — so the same gap yields z −15.9 instead of z −7.6. The degree-preserving rewire (test D, 4.6 pp) is the control whose numbers the text quotes.

**Figure 9.** The proteinopathy proteins are poised to damage the synapse by sequestering 44 proteins essential for synaptic structure and functions. Readout in A–C: the 37 SynGO-significant hub proteins that APP, HTT, SNCA and MAPT do not bind, across 154 HPA cell types, 25 draws per control.

(A) Of the 81 SynGO-significant proteins among the 351 hub proteins, 44 are themselves proteinopathy targets against a degree-matched expectation of 24.1 ($z = +5.73$), leaving 37 bystanders.

(B) Cascade retained by the 37 bystanders, as a percentage of the same cell type unperturbed, under a degree-matched withdrawal, the observed withdrawal of the real 128, and a per-cell-type degree-matched non-synaptic bystander control. Medians: proteins reached 98.5 / 91.7 / 91.6; dimer pairs 92.0 / 78.4 / 76.3; limiting capacity 91.9 / 72.4 / 72.3. The real target set removes 21.6% of the bystanders' dimer pairs against 8.0% for a degree-matched withdrawal, in 154/154 cell types; the non-synaptic control is null (capacity $z = 0.00$).

(C) Excess damage per cell type, z against the degree-matched withdrawal, by HPA class. The spread is 4.0× a leave-one-out null, but neuronal cell types are not its extreme (neuronal against rest $p = 0.52$); germ, muscle and squamous epithelial classes carry the excess while neurons carry the largest absolute loss.

(D) Median share of the lost dimer pairs returned by freeing one target, 124 live targets over 20 CNS-relevant cell types; best single 7.7%, best ten 35.6%.

(E) Median capacity share per target under raw nTPM against rank-transformed abundance. Neutralising expression moves CALM1 from rank 1 to 7 and CALM3 from 3 to 27; MAPK1 rises to 1.

(F) The 124 single-protein recoveries sum to 124.8% of the loss on dimer pairs, 115.4% on raw-abundance capacity and 102.3% once abundance is rank-transformed, against the 100% that full relief returns by definition — near additivity, so partial clearance buys roughly its share and nothing has to be cleared first. The synapse is not itself fragile, freeing synaptic targets one at a time is close to additive, and freeing the individual targets recovers the lost protein dimers.

**Table 1. Cascade and assembly in two cellular proteomes.**

| | Brain excitatory neurons | Hepatocytes |
|---|---|---|
| **Available proteins** | 7,997 | 6,888 |
| **Panel proteins available (of 1,122)** | 696 | 612 |
| **Initiator TFs available (of 716)** | 421 | 348 |
| **TF–TF seeds available / plateau** | 300 / 150 | 277 / 153 |
| **HUB seeds available / plateau** | 290 / 216 | 264 / 211 |
| **Proteins engaged (union)** | 339 | 337 |
| **Homodimer-capable proteins** | 171 | 171 |
| **Mutually engaged heterodimers** | 5,923 | 6,000 |
| **Dimers inside the 1,122-protein panel** | 7,694 | 7,610 |
| **CORUM complexes at q < 0.05** | 257 | 225 |
| **Complexes with every subunit engaged** | 28 | 25 |

Legend: Cutoff expression of cellular proteomes was ($nTPM \geq 1$).

**Table 2.** Identification of 128 binding targets of proteinopathy proteins among 1,122 PPI cascade constituents.

| Experimental category | Principal initiator TFs (n=20) | | Complementary initiator TFs (n=5) | | TF-TF PPI members (n=40) | | TF-HUB protein members (n=97) | |
|---|---|---|---|---|---|---|---|---|
| **Proteinopathy proteins** | **TF-TF PPI cascade branch** | **TF-HUB PPI cascade branch** | **TF-TF PPI cascade branch** | **TF-HUB PPI cascade branch** | **TF-TF PPI cascade branch** | **TF-HUB PPI cascade branch** | **TF-TF PPI cascade branch** | **TF-HUB PPI cascade branch** |
| **APP** | 14 | 1 | 5 | 1 | 28 | 2 | 5 | 35 |
| **HTT** | 6 | 3 | 0 | 0 | 12 | 5 | 6 | 25 |
| **SNCA** | 1 | 0 | 0 | 0 | 3 | 1 | 2 | 40 |
| **MAPT** | 0 | 0 | 0 | 0 | 1 | 0 | 1 | 46 |

**Table 3.** The list of winners of the therapeutic candidate contest in the clinical degeneration proteinopathy proteins co-pathology settings

| Arm | Readout | 1st | 2nd | 3rd | top-10 share | median target |
|---|---|---|---|---|---|---|
| **SNCA+MAPT** | dimer pairs | **MAPK1 9.864%** | CALM1 9.111% | PPP2CA 8.706% | 70.8% | 1.409% |
| **SNCA+MAPT** | capacity | **MAPK1 15.423%** | CALM1 14.407% | CALM3 12.771% | 74.3% | 1.111% |
| **APP+MAPT** | dimer pairs | **SRC 9.712%** | YWHAB 9.216% | MAPK1 8.772% | 86.9% | 1.103% |
| **APP+MAPT** | capacity | **MAPK1 10.182%** | CALM1 7.933% | CALM3 7.098% | 64.5% | 0.710% |

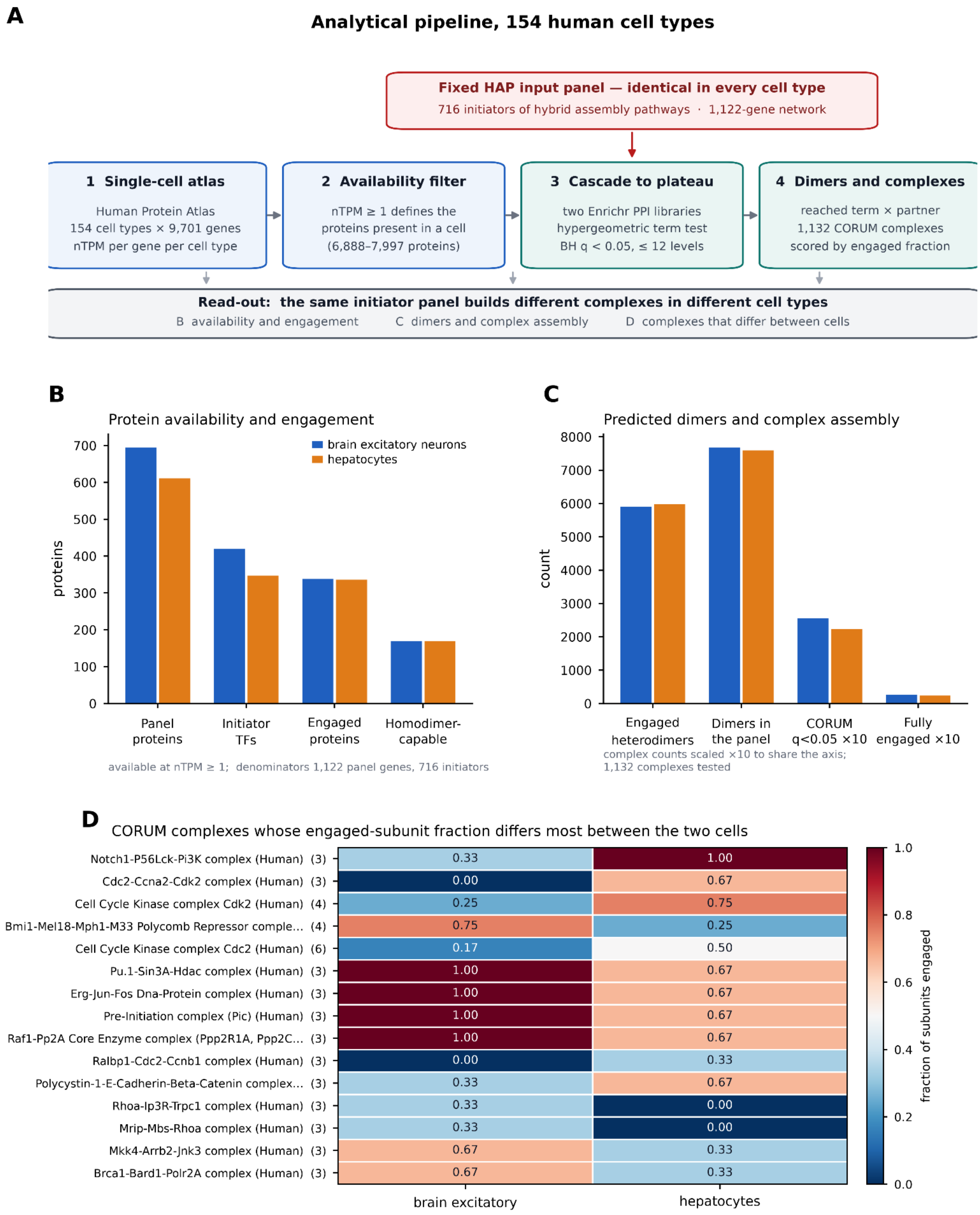

D shows the top 15 complexes by absolute difference; the number of annotated subunits is given in brackets. Availability is nTPM ≥ 1 in the Human Protein Atlas single-cell consensus. The cascade is seeded with the initiators available in that cell and iterated to its plateau; a protein is engaged when it is a cascade product, and a dimer is mutually engaged when both partners are. Complex enrichment is hypergeometric against that cell's own available proteome.

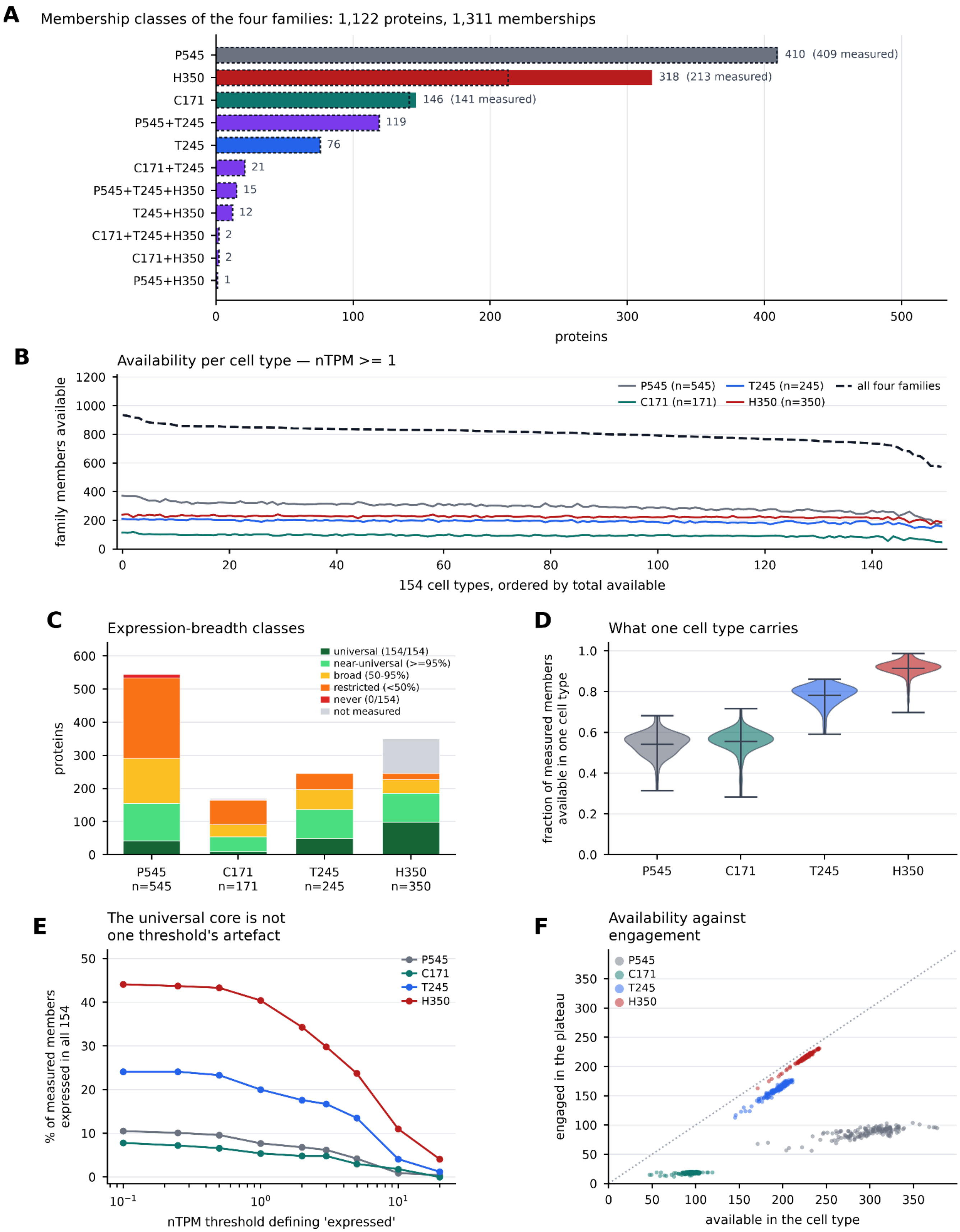

All panels use the nTPM ≥ 1 availability rule on the Human Protein Atlas single-cell consensus, 154 cell types × 9,701 genes. In A the filled bar is the class size and the dashed outline is how much of it the atlas measures; 105 H350 members are absent from the atlas and are carried as 'not measured' throughout. Breadth classes in C are disjoint within a family. In F a point is one family in one cell type, and the dotted line is engagement equal to availability.

F2

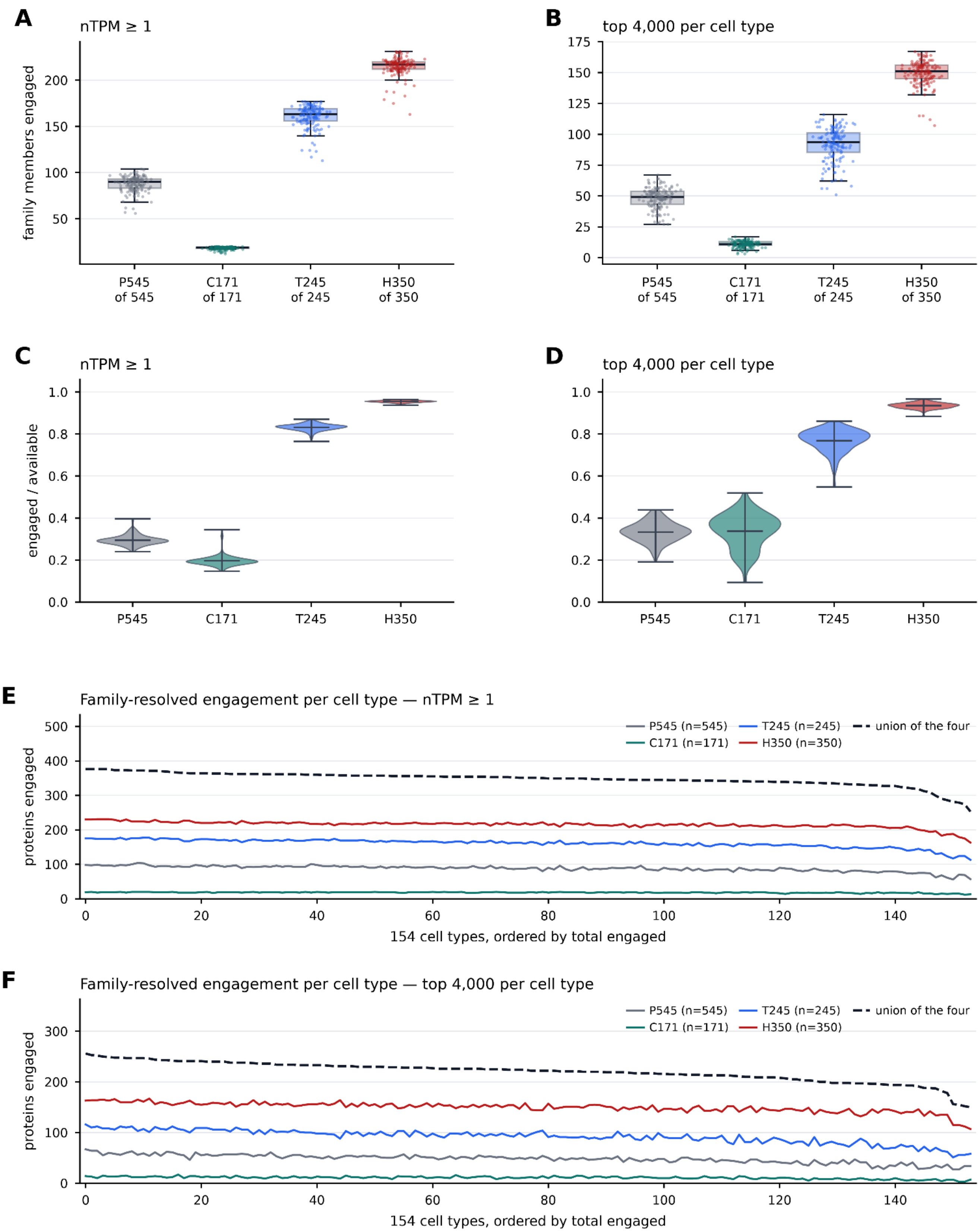

A protein is engaged when the initiator-seeded cascade reaches it at its plateau. The two rules differ only in what counts as available: nTPM ≥ 1 in the Human Protein Atlas single-cell consensus (A, C, E), or membership of that cell type's 4,000 highest-nTPM genes (B, D, F). Denominators in A and B are whole family sizes; 105 of the 350 H350 members are absent from the atlas and can never be engaged. C and D divide by what is available in that cell type instead.

**F3**

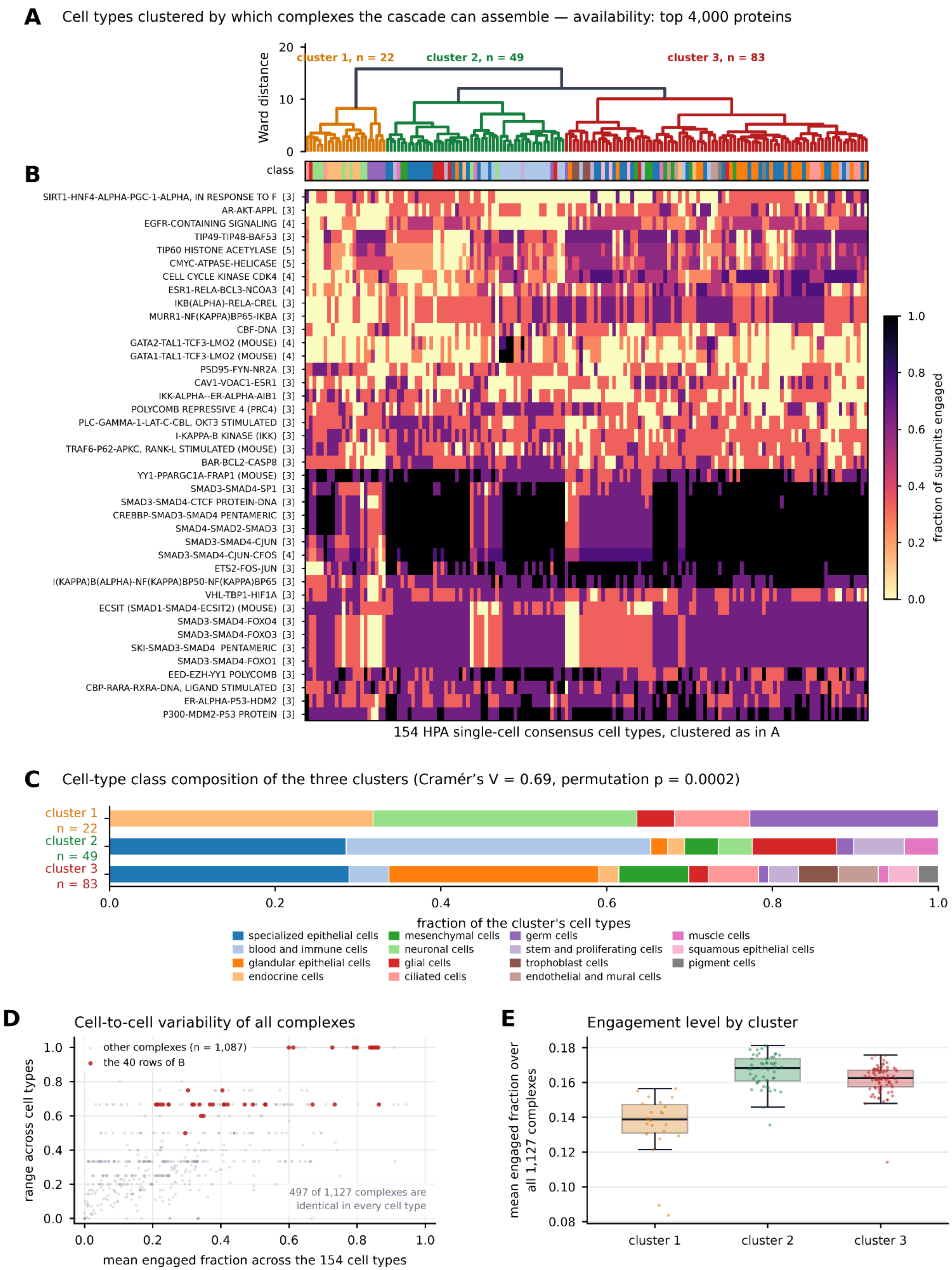

Availability is membership of a cell type's 4,000 highest-nTPM genes; a complex's engaged fraction is the share of its annotated subunits the initiator-seeded cascade reaches at its plateau, over the 1,127 CORUM complexes with at least three subunits. The linkage in A uses the 200 most cell-variable complexes and is cut at 0.7 × its maximum height. Cell-type names are omitted from A and B for legibility and are given in tables/figure4_cell_clusters.csv.

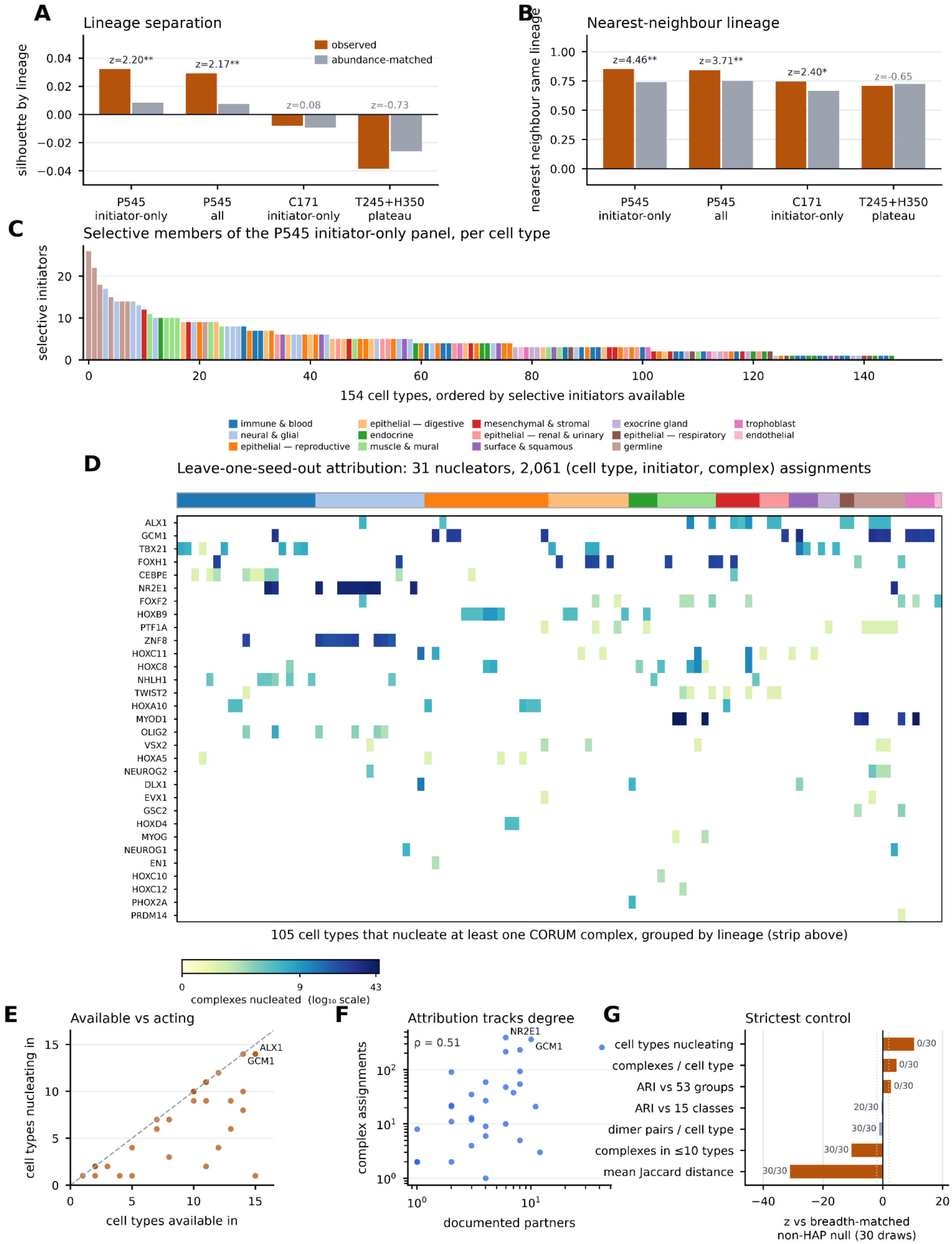

Nucleation is direct-partner assembly with no enrichment step at the seed end, attributed by leaving one seed out at a time. Orange in G marks a metric the observed seeds clear at |z| > 2 in the favourable direction; the labels give draws at or above observed. Observed mean pairwise Jaccard distance is 0.810 (cophenetic 0.970). Recovery of the 15 HPA classes is the one metric that does not survive matching on availability breadth — 20 of 30 matched non-HAP draws exceed the observed ARI of 0.0946 — so class-level structure follows from cell-type restriction, not from HAP membership. ARI is scale-dependent: only the same-cut null comparison is meaningful. An adult atlas cannot test direction.

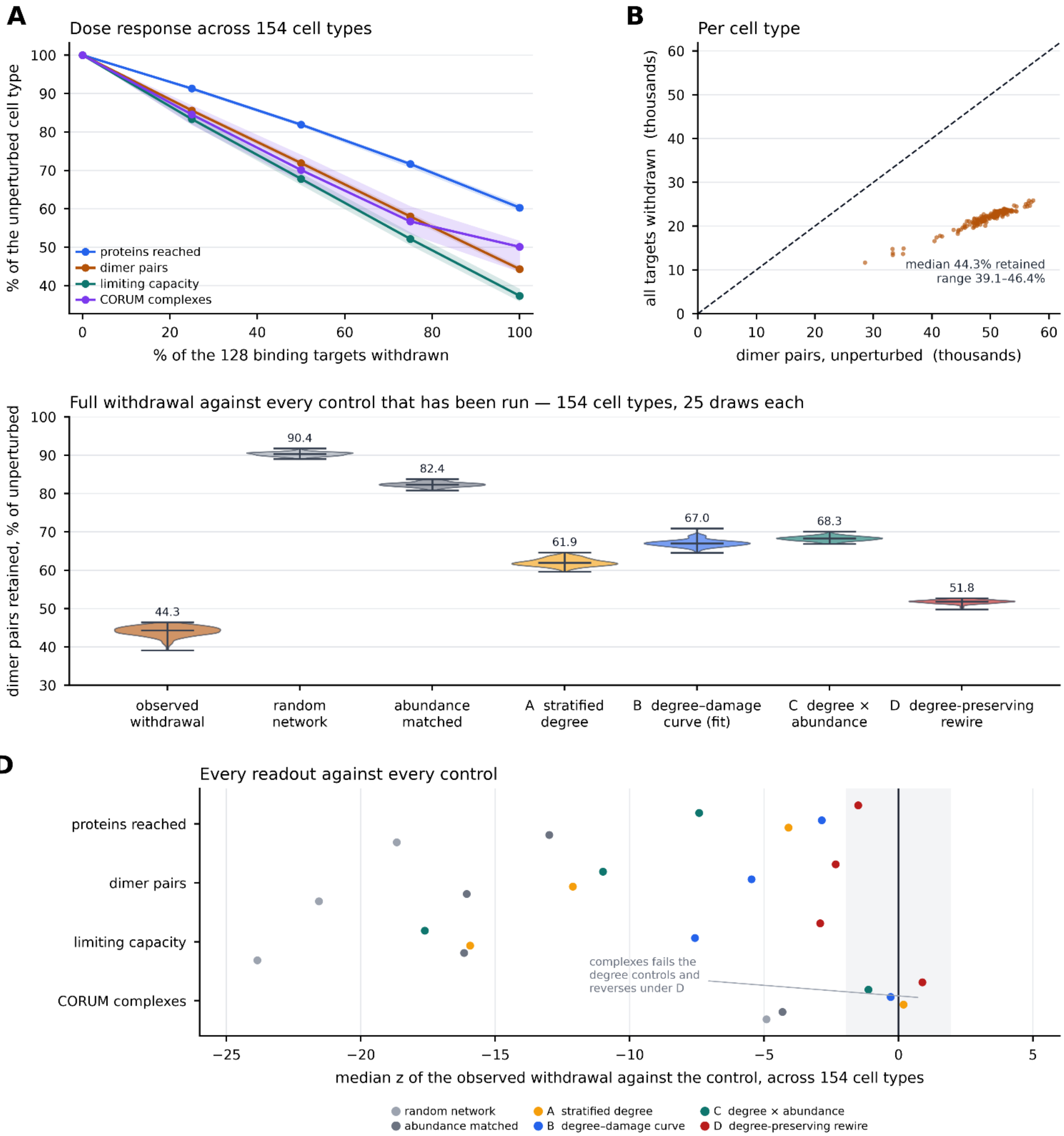

Binding is modelled as complete, permanent withdrawal from the cell type's available pool, so these are ceilings on the effect rather than expected sizes. The 128 targets are the network's connectors (median documented degree 53 against 9 for the rest of the 1,122-protein network; median nTPM 68.7 against 15.5), which is why the degree controls in C and D carry the argument and the size-matched ones do not. The four readouts come from one pass over the same reached set and are not independent. CORUM complexes is a thresholded enrichment call: it sits where connectivity predicts under the curve ($z = -0.30$) and reverses under the rewire ($z = +0.88$, worse than every draw in 0 of 154 cell types), and must be reported as a failed readout. No dose corresponds to a known disease state, and nothing here shows the cascade stalls because of binding rather than through shared hub identity.

F6

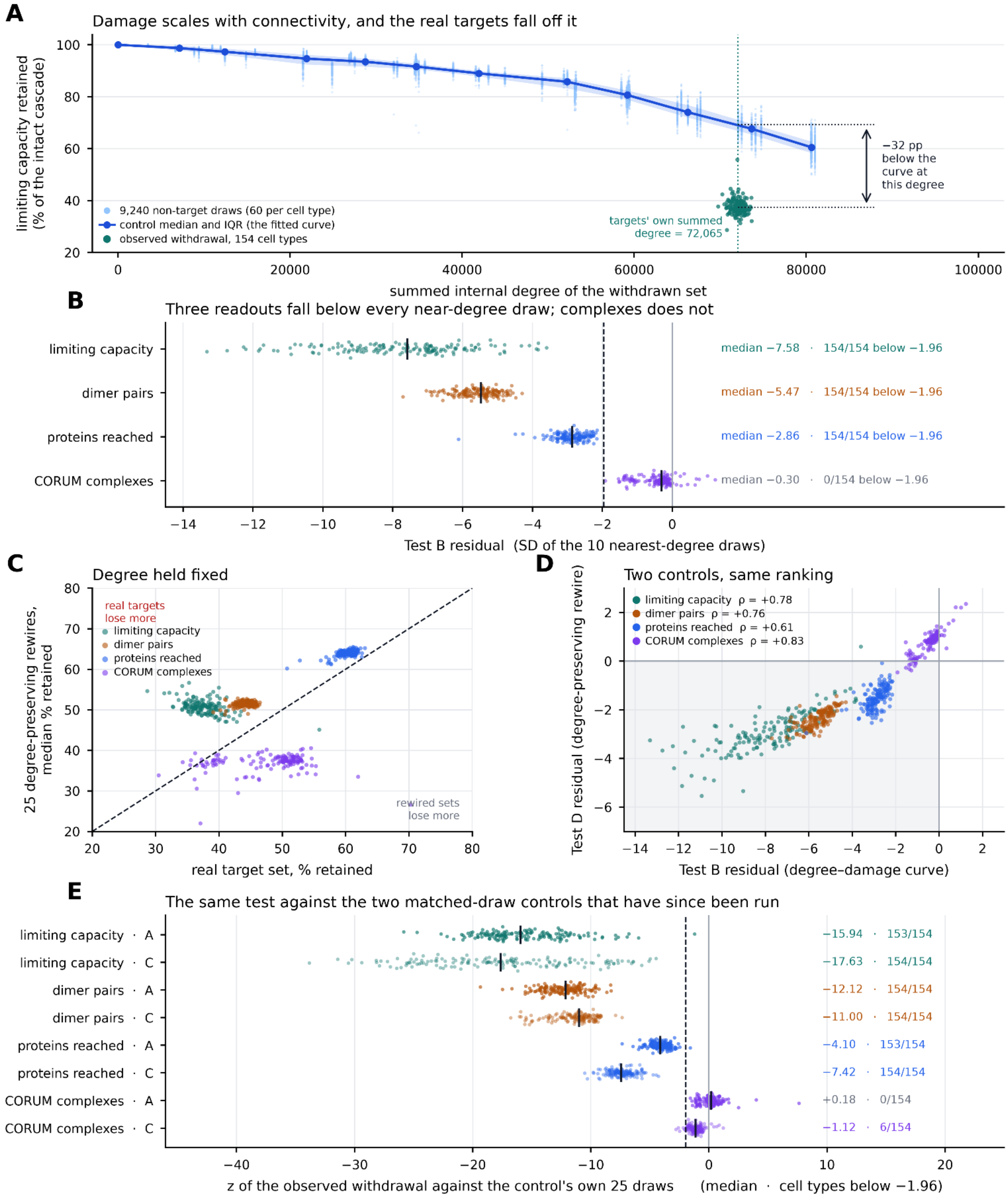

Every point is one of the 154 HPA cell types; 128 terms are withdrawn in each. The four readouts come from one pass over the same reached set — limiting capacity is dimer pairs weighted by min(nTPM) over the same partner pairs — so their agreement is not independent replication. A and C resample which terms are withdrawn and leave the library's structure intact; D preserves every protein's degree and every term's size exactly, which is why its z values are the smallest and the ones to quote. Test C withdraws 122 terms rather than 128 (6 short in the highest degree × highest abundance cell of its grid, where the targets themselves sit), so its gap is an upper bound. CORUM complexes clears no control: +0.18 under A, −0.30 under B, −1.12 under C, and it reverses under D (+0.88, worse than every draw in 0 of 154 cell types). It is reported here as a failed readout and supports nothing in either direction.

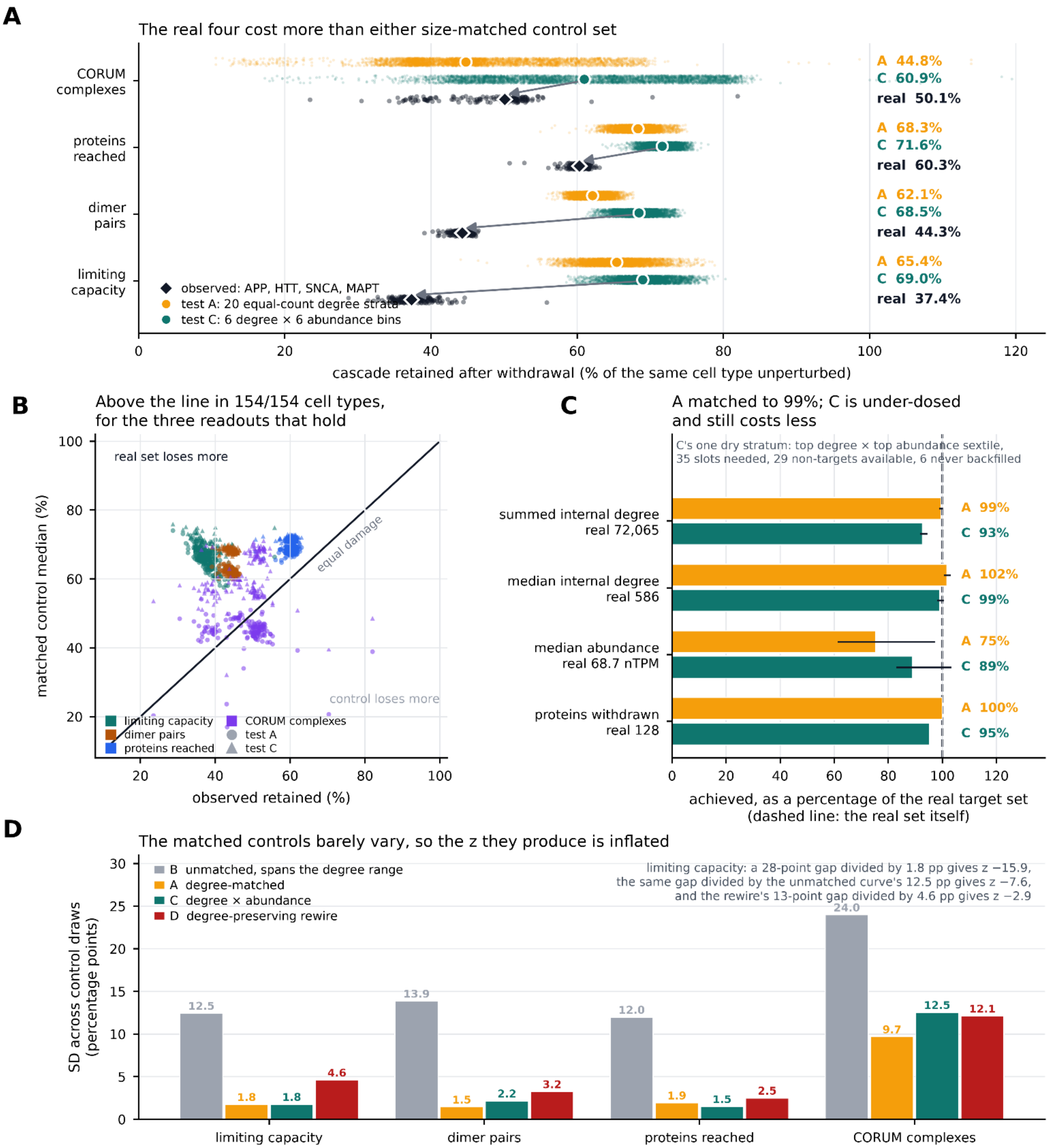

Every point is one of the 154 HPA cell types; 128 network terms are withdrawn in each, 25 draws per test. The four readouts come from one pass over the same reached set — limiting capacity is dimer pairs weighted by min(nTPM) over the same partner pairs — so their agreement is not independent replication. CORUM complexes fails both matched controls (z +0.18 under A, −1.12 under C) and is drawn here alongside the three that do not; it supports nothing in either direction. Panels A, C and D pool all 3,850 draws per test; panel B takes each cell type's own control median. A larger control SD is the honest denominator: matching the propagation variable this tightly collapses it and inflates z without the gap itself changing, which is why the rewire's numbers are the ones quoted in the text.

F8

## Withdrawal of the proteinopathy-bound cascade constituents, and what returning them recovers

Readout throughout: the 37 synaptic proteins the proteinopathy proteins do not bind, across the 154-cell-type atlas

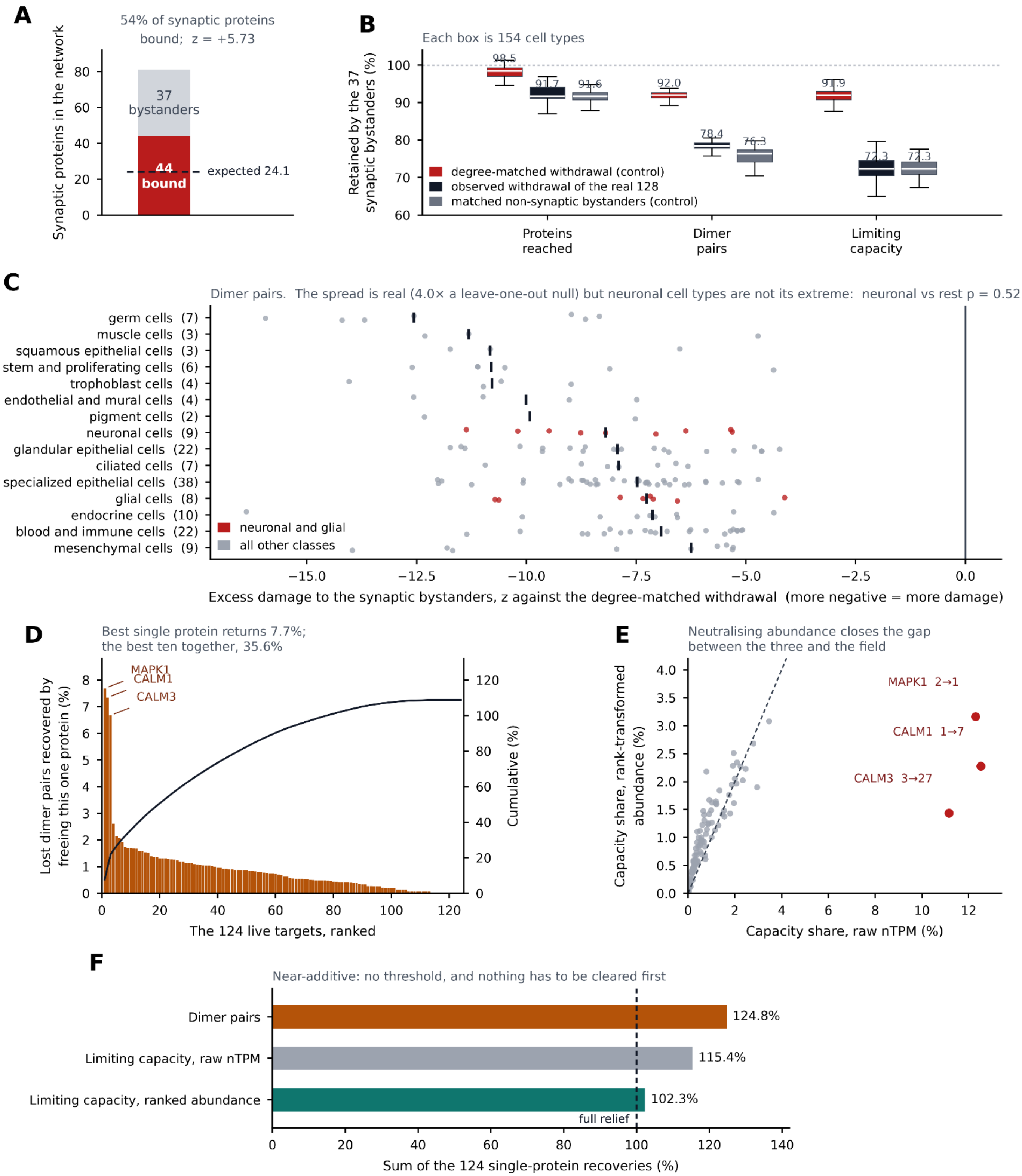

The 128 cascade constituents bound by APP, HTT, SNCA and MAPT are withdrawn from the availability set and the cascade re-run to its plateau in each of the 154 cell types. Synaptic proteins are the 81 network members covered by the 23 SynGO terms significant at BH $q < 0.05$ over the 351 hub proteins; the 44 that are themselves targets are excluded from every readout, leaving 37 bystanders. The degree-matched control in B and C draws 25 replacement sets stratified on cascade-internal degree with the bystanders held out of the pool, so it cannot remove the thing being measured; each draw is 123 rather than 128 proteins and carries 68,060 against 72,065 summed internal degree, a 5.6% deficit that biases it toward less damage. Baselines are taken over the members surviving each run. D to F are the 20 CNS-relevant cell types only, at 124 targets live in at least one of them. The CORUM complexes readout is omitted: it fails every degree control and reverses under test D.